\documentclass{article}
\usepackage{url}
\usepackage[dvips]{graphicx}
\usepackage{natbib}
\usepackage{subfigure}
\usepackage{fancybox}
\usepackage{calrsfs}
\usepackage{color}
\usepackage{amsmath}
\usepackage{amsthm}
\newcommand{\rs}[1]{\mathrsfs{#1}}
\newcommand{\eqnum}[1]{(\ref{#1})}
\newcommand{\thatis}{{i.e.}}
\newcommand{\argmin}{\mathop{\mathrm{argmin}}}

\newcommand{\sbar}{\,|\,}
\newcommand{\trace}{{\rm Tr}}

\newcommand{\Remp}{L_D}
\newcommand{\loss}{L}
\newcommand{\risk}{R^*}
\newcommand{\riskhat}{\widehat{R}_D}
\newcommand{\Var}{{\rm Var}}

\newtheorem{theorem}{\indent Theorem}

\newtheorem{definition}[theorem]{\indent Definition}

\title{A Note on Model Selection \\for Small Sample Regression}
\author{Masanori Kawakita and Jun'ichi Takeuchi\\
the faculty of Information Science and Electrical Engineering\\
Kyushu University\\ 744 Motooka, Nishi-Ku, Fukuoka City, Fukuoka 819--0395, Japan\\
}
\date{}
\begin{document}
\maketitle
\begin{abstract}
The risk estimator called ``Direct
 Eigenvalue Estimator'' (DEE) is studied. DEE was developed for small
 sample regression. 
 In contrast to many existing model selection criteria, derivation of
 DEE requires neither any asymptotic assumption nor any prior knowledge
 about the noise variance and the noise distribution. 
It was reported that DEE performed well in small sample
 cases but DEE performed a little worse than the state-of-the-art ADJ. 
This seems somewhat counter-intuitive because DEE was developed for
 specifically regression problem by exploiting available information
 exhaustively, while ADJ was developed for general setting.
 In this paper, we point out that the derivation of DEE includes an
 inappropriate part in spite that the resultant form of DEE is valid in
 a sense. As its result, DEE cannot derive its potential.
We introduce a class of `valid' risk estimators based on the idea of DEE
 and show that better risk estimators (mDEE) can be found in the class. 
By numerical experiments, we verify that mDEE often performs better than
 or at least equally the original DEE and ADJ. 
 \end{abstract}
\section{Introduction}
The most famous approach of model selection is to derive a
risk estimator and to choose the model minimizing it.  This type of
model selection includes cross-validation and so-called ``information
criteria'' including AIC \citep{akaike72}, BIC \citep{schwartz79}, GIC
\citep{konishikitagawa96} and so on. Basically, information criteria
have been derived by using asymptotic expansion, which requires the
sample number $n$ to go to the infinity.  Though the cross-validation was not derived by
asymptotic assumption, its unbiasedness or performance is guaranteed
basically by asymptotic theory.
For regression, \citet{chapelleetal02} proposed an
interesting model selection criteria called Direct
Eigenvalue Estimator (DEE). DEE has the following remarkable
characteristics: (i) DEE is an approximately unbiased risk estimator for
finite $n$.  (ii) no asymptotic assumption ($n\rightarrow \infty$) is necessary to derive
it.  (iii) no prior knowledge about the noise variance and the noise
distribution is necessary. Due to these virtues, DEE is expected to
perform nicely for small sample cases. DEE requires two assumptions in
its derivation instead of using the asymptotic assumption. The first
assumption is that a large number of unlabeled data are available in
addition to the labeled data. This assumption is often made in recent
machine learning literatures and is practical because the unlabeled data
can usually be gathered automatically. 
The other assumption is most important, which 
imposes statistical independence between the inside-the-model part and the
outside-the-model part of the dependent variable $y$. 
This assumption holds not exactly in general but hold approximately. By
numerical experiments, \citet{chapelleetal02} showed that DEE performed
better than many of the conventional information criteria and the
cross-validation. However, they also reported that another criterion ADJ
\citep{schuurmans97} often performed better than DEE.  It should be noted here
that the comparison between ADJ and DEE is fair since ADJ also assumes
that a lot of unlabeled data are available. Even though ADJ is the
state-of-the-art, that result seems somewhat strange because DEE was
derived specifically for regression by exploiting the properties of
regression exhaustively while the derivation of ADJ is somewhat
heuristic and was developed for general setting.
By careful investigation, we found an inappropriate part in the
derivation process of DEE, although the resultant form of DEE is `valid'
in a sense. As a result, DEE cannot derive its potential.
To clarify these facts, we formulate the derivation process of DEE again
and introduce a class of `valid' risk estimators based on the idea of 
DEE. Then, we show that DEE belongs to this class but is not close to
the optimal estimator among this class. Indeed, we can find several more
reasonable risk estimators (mDEE) in this class. 
 The variations arise from how to balance a certain bias-variance
 trade-off. The performance of mDEE is investigated by
 numerical experiments. We compare the performance of mDEE with the
 original DEE, ADJ and other existing model selection methods.
 
This paper is an extended version of the conference paper
 \citep{kawakitaetal10}. We pointed out the above inappropriate part in
 the derivation of DEE and proposed
 a naive modification in the paper. However, theoretical analysis and numerical experiments are
 significantly strengthened in this paper.
 
The paper is organized as follows. We set up a regression problem and
introduce some notations in Section \ref{regression}. In Section
\ref{chapelle}, we briefly review the result of \citet{chapelleetal02}
and explain which part is inappropriate in the derivation of DEE.
In Section \ref{proposal}, a class of valid risk estimators are
defined. We explain why DEE is valid but not close to the optimal
estimator in this class. In addition, some modifications of DEE are
proposed. Section \ref{sim} provided numerical experiments to
investigate the performance of our proposal. The conclusion is described in Section \ref{conclusion}.

\section{Setup and Notations}\label{regression}
We employ a usual regression setting as reviewed briefly below.  
Let $\rs{X}\subset \Re^M$ and $\rs{Y}:=\Re$. 
Suppose that we have training data $D:=\{(x_1,y_1), (x_2,y_2),\cdots, (x_n,y_n)\}$ generated from the joint density $p(x,y)=p(x)p(y|x)$
i.i.d.\ (independently and identically distributed), where $(x_i,y_i)\in
\rs{X}\times \rs{Y}$ for each $i$. 
Here, we further assume the following regression model:
\begin{equation}
y_i=f_*(x_i)+\xi_i, \label{regmodel}
\end{equation}
where $f_*(x)$ is a certain regression function belonging to
$L^2(\rs{X})$ and $\xi_i$ is a noise random variable which is subject to
$p_{\xi}(\xi)$ with mean zero and variance 
$\sigma^2$ and is independent of $x$.  
This implies that $p(y|x)=p_{\xi}(y-f_*(x))$.
The goal of regression problem is to estimate $f_*(x)$ based on the
given data set.
To estimate $f_*(x)$, let us consider a model of regression function defined by
\begin{equation}
 f(x;\bar{\alpha}):=\sum_{k=1}^{\infty}\bar{\alpha}_k\phi_k(x),
\end{equation}
 where $\bar{\alpha}=(\bar{\alpha}_1,\bar{\alpha}_2,\cdots
 )^T$. Here, $T$ denotes the transposition of vectors or a matrices. Just
 for convenience, we can assume that $\{\phi_k(x)\}$ is a basis of 
 $L_2(\rs{X})$. If not, we can always extend
 it as such without loss of generality. By this
 assumption, there exists $\bar{\alpha}^*:=(\bar{\alpha}^*_1,\bar{\alpha}^*_2,\cdots)^T$
 such that $f_*(x)\equiv f(x;\bar{\alpha}^*)$ almost
 everywhere.
Our task now reduces to find an estimator $\hat{f}(x)$ of $f(x;\bar{\alpha}^*)$ as accurate 
 as possible. Its accuracy is measured by the loss function (Mean Squared Error) defined as
\begin{equation}
 \loss(\hat{f}):=E_{x,y}[(y-\hat{f}(x))^2]. \label{loss}
\end{equation}
Here, $E_{x,y}[\cdot ]$ denotes the
 expectation with respect to random variables $x,y$. Similarly,
 each expectation $E$ has subscripts expressing over which random
 variables the expectation is taken. Since
 $f(x;\bar{\alpha})$ essentially can express an arbitrary element of 
$L^2(\rs{X})$, $f(x;\bar{\alpha})$ itself is too flexible and tends to
cause overfitting in general. Therefore, we usually use a truncated
version of $f(x;\bar{\alpha})$ as a model 
\[
 \rs{M}_d:=\bigg\{f_d(x;\alpha):=\sum_{k=1}^d\alpha_k\phi_k(x)\bigg|
 \alpha=(\alpha_1,\alpha_2,\cdots, \alpha_d)^T\in \Re^d\bigg\},
\]
where $d$ is a positive integer. 
The ideal estimate of parameter $\alpha$ is obtained by minimizing the
loss function $\loss $ in \eqnum{loss} with respect to
$\alpha$. However, $\loss$ is not available because the distribution
$p(x,y)$ is unknown. Therefore, we usually minimize an empirical loss function based on $D$, which is defined as
\begin{equation}
 \Remp(f_d(\cdot;\alpha)):=\frac{1}{n}\sum_{i=1}^n(y_i-f_d(x_i;\alpha))^2.
\end{equation}
For notation simplicity, we define $y=(y_1,y_2,\cdots, y_n)^T$ and
$\Phi$ as $(n\times d)$ matrix whose $(i,k)$ component is
$\phi_k(x_i)$. Then, $\Remp(f_d(\cdot;\alpha))=(1/n)\|y-\Phi\alpha\|^2$,
where $\|\cdot \|$ is the Euclidean norm. 
The estimator $\hat{\alpha}(D)$ minimizing $\Remp$ is referred to as Least
Squares Estimator (LSE), \thatis,  
\begin{equation}
 \hat{\alpha}(D):=\argmin_{\alpha\in \Re^d}\Remp(f_d(\cdot;\alpha))=(\Phi^T\Phi)^{-1}\Phi^Ty. 
\end{equation}
We sometimes drop $(D)$ of $\hat{\alpha}(D)$ for notation simplicity. 
An important task is to choose the optimal $d$.
When $d$ is too large, $f_d$ tends to overfit, while $f_d$ cannot
approximate $f_*(x)$ with too small $d$.
To choose $d$, we assume that additional unlabeled data
$D_U:=\{x'_1,x'_2,\cdots,x'_{n'}\}$ are available, where each $x'_j$ is
subject to $p(x)$ independently.
The number of unlabeled data $n'$ is assumed to be significantly larger than $n$. 
Note that $D_U$ is used not for
parameter estimation but only for model selection as well as
\citet{chapelleetal02}. 
The basic idea to choose $d$ is as follows. The following risk (expected
loss function)
\begin{equation}
 \risk(d):=E_D[\loss(f_d(\cdot;\hat{\alpha}(D)))]
\end{equation}
is often employed to measure the performance of the model. Hence, one of
natural strategies is deriving an estimate of $\risk(d)$ (denoted by
$\riskhat(d)$) using $D\cup D_U$, and then choosing the model as
$\hat{d}:=\argmin_{d}\riskhat(d)$. 
Many researchers have proposed estimators of $\risk(d)$ so far. 
In the next section, we introduce one of such estimators,  which was
proposed by \citet{chapelleetal02}. 

\section{Review of Direct Eigenvalue Estimator}\label{chapelle}
Most of past information criteria have been derived based on asymptotic
expansion. That is, they postulate that $n\rightarrow \infty$. 
In contrast, \citet{chapelleetal02} derived a risk estimator called DEE
(Direct Eigenvalue Estimator) without using asymptotic assumption.  
In this section, we briefly review DEE and explain that its derivation
includes an inappropriate part. 
As is well known, $\Remp(f_d(\cdot;\hat{\alpha}(D)))$ is not an
unbiased estimator of $\risk (d)$. That is, 
\[
\mbox{bias}(\Remp(f_d(\cdot;\hat{\alpha}(D)))):=E_D[\Remp(f_d(\cdot;\hat{\alpha}(D)))-\risk (d)]
\]
is not equal to zero. 
Let $T^*(n,d)$ be
\[
 T^*(n,d)=\frac{E_D[\loss(f_d(\cdot;\hat{\alpha}(D)))]}{E_D[\Remp(f_d(\cdot;\hat{\alpha}(D)))]}.
\]
Using $T^*(n,d)$, let us consider the following risk estimator
\[
\riskhat^*(d):=T^*(n,d)\Remp(f_d(\cdot;\hat{\alpha}(D))).
\]
It is immediate to see that this estimator is exactly unbiased, \thatis,
$\mbox{bias}(\riskhat^*(d))=0$. That is, $T^*(n,d)$ is a so-called bias-correcting term. 
Remarkably, this estimator corrects the bias multiplicatively, whereas the
most of existing information criteria correct the bias additively like AIC. 
\citet{chapelleetal02} showed that $T^*(n,d)$ can be calculated as
the following theorem. 
   \begin{theorem}[\citealp{chapelleetal02}]\label{DEEderivation}
    Let $\phi_1(x)\equiv 1$. Define $\widehat{C}:=(1/n)\Phi^T\Phi$.
    Suppose that the following assumptions hold.n
   \begin{itemize}
    \item[A1] Assume that $\{\phi_k(x)|k=1,2,\cdots, d\}$ is orthonormal with
	      respect to the expectation inner product
	      $<a(x),b(x)>_p:=E_x[a(x)b(x)]$, \thatis,   
 \begin{equation}
\forall k,\,\forall k',\quad <\phi_k(X),\phi_{k'}(X)>_p=\delta_{kk'},\label{orthassumption}
 \end{equation}
 where $\delta_{kk'}$ is Kronecker's delta.
    \item[A2] Let $f^*_d:=\min_{f\in
		 \rs{M}_d}\loss(f)$. Define $\tilde{y}_i:=y_i-f^*_d(x_i)$ and
   $\tilde{y}:=(\tilde{y}_1,\tilde{y}_2,\cdots, \tilde{y}_n)$. Assume
	      that
	      \begin{equation}
 \mbox{``}\tilde{y} \mbox{ and } \Phi \mbox{ are statistically independent.''}\label{siassumption}
	      \end{equation}  
   \end{itemize}
Then $T^*(n,d)$ is exactly calculated as
 \begin{equation}
 T^*(n,d)=\frac{1+(1/n)E_D[\trace(\widehat{C}^{-1})]}{1-(d/n)}. \label{predee}
 \end{equation}
 \end{theorem}
See \citet{chapelleetal02} for the meaning of the assumption A2. A key
 fact is that Theorem \ref{DEEderivation} holds for finite $n$ and the
 resultant form of $T^*(n,d)$ does not depend on any unknown quantities
 except $E_D[\trace(\widehat{C}^{-1})]$.
 In addition, it seems to be not difficult to find valid estimators
 of $E_D[\trace(\widehat{C}^{-1})]$. Indeed, \cite{chapelleetal02} derived its estimator as
\begin{equation}
\trace\left(\widehat{C}^{-1}\widetilde{C}\right), \label{deepestimator}
\end{equation}
where $\widetilde{C}$ is defined by
$\widetilde{C}=\frac{1}{n'}(\Phi')^T\Phi'$ and $\Phi'$ is an
$(n'\times d)$ matrix whose $(j,k)$ component is $\phi_k(x'_j)$.
The resultant risk estimator is called DEE and is given by 
\begin{equation}
 \riskhat^{\rm DEE}(d)=\frac{1+(1/n)\trace(\widehat{C}^{-1}\widetilde{C})}{1-(d/n)}\Remp(f_d(\cdot;\hat{\alpha}(D))). \label{finaldee}
\end{equation}
Note that the resultant bias correction factor is invariant under coordinate
transformation. That is, the orthonormal assumption
(\ref{orthassumption}) can be removed by chance. However, this is
somewhat queer. DEE was derived based on (\ref{predee}) 
but (\ref{predee}) is not invariant under coordinate transformation (it was
derived by assuming the orthonormality of basis).
Actually, (\ref{deepestimator}) is not a consistent estimator of
$E_D[\trace(\widehat{C}^{-1})]$ in non-orthonormal case.
This is because the derivation of estimator (\ref{deepestimator})
includes an inappropriate part. We explain it in the remark at the end
of this section. 
However, we must emphasize that the resultant form of DEE in (\ref{finaldee})
is valid as a risk estimator in a sense in spite of the above fact. 
Indeed, DEE dominated other model selection methods in numerical experiments, as
reported by \citet{chapelleetal02}. 
However, due to the inappropriate derivation, DEE cannot demonstrate
its potential performance. We will explain it in details in the next section. 
\paragraph{Remark}
We explain here which part of the derivation of DEE is
 inappropriate. \cite{chapelleetal02} derived (\ref{deepestimator}) as follows.  
First, they rewrote $E_D[\trace(\widehat{C}^{-1})]$ as
\[
 E_D[\trace(\widehat{C}^{-1})]=E_D\left[\sum_{k=1}^d(1/\lambda_k)\right],
\]
where $\lambda_k$ is the $k$-th eigenvalue of $\widehat{C}$.
The subsequent part is described by quoting the corresponding part of their
paper (page 16 of \citet{chapelleetal02}). Note that some
notations and equation numbers are replaced in order to be consistent
with this paper.
\newtheorem{theo}{Quote}{}
\begin{center}
\begin{theo}[Derivation of DEE]
In the case when along with training data,
“unlabeled” data are available (x without y), one can compute two covariance matrices:
one from unlabeled data $\widetilde{C}$ and another from the training data $\widehat{C}$.
 There is a unique matrix $P$ (\citealp{hornjohnson85}; Corollary 7.6.5) such that
\begin{equation}
 P^T\widetilde{C}P =I_d \mbox{ and } P^T\widehat{C}P=\Lambda, \label{critical}
\end{equation}
where $\Lambda$ is a diagonal matrix {\bf with diagonal elements
 $\bf \lambda_1,\lambda_2,\cdots, \lambda_d$}. To perform model selection, we
 used the correcting term in \eqnum{predee}, where we replace $E\sum_{k=1}^d1/\lambda_k$ with its
 empirical
value,
\begin{eqnarray}
  \sum_{k=1}^d1/\lambda_k&=&{\rm Tr}\left(P^{-1}\widehat{C}^{-1}(P^T)^{-1}P^T\widetilde{C}P\right)\nonumber \\
&=&{\rm Tr}\left(\widehat{C}^{-1}\widetilde{C}\right). \label{deederivation}
\end{eqnarray}
\end{theo}
\end{center}
However, Corollary 7.6.5 in \citet{hornjohnson85} does not
guarantee the existence of a matrix $P$ satisfying \eqnum{critical}. We
quote the statement of the corollary. 
\begin{theo}[Corollary 7.6.5 in \citet{hornjohnson85}]
If $A\in M_n$ is positive definite and $B\in M_n$ is Hermitian, then
 there exists a nonsingular matrix $C\in M_n$ such that $C^*BC$ is
 diagonal and $C^*AC=I$.
\end{theo}
Here, $M_n$ denotes the set of all square matrices of
dimension $n$, whose elements are complex numbers. Furthermore, the
symbol $*$ denotes the Hermitian adjoint. As seen in the quote, the above corollary only guarantees that
$P^T\widetilde{C}P=I_d$ and $P^T\widehat{C}P$ is diagonal.
However, Quote 1 claims that $P^T\widehat{C}P$ must be not only a
diagonal matrix but also a diagonal matrix whose elements are equal to
the eigenvalues of $\widehat{C}$ (See the bold part of Quote 1). 
This claim does not hold correct in general.
Indeed, \eqnum{deederivation} is queer. Since \eqnum{deederivation}
implies that $\trace(\widehat{C}^{-1})=\trace(\widehat{C}^{-1}\widetilde{C})$ for any
unlabeled data, unlabeled data plays exactly no role.
\section{Modification of Direct Eigenvalue Estimator}\label{proposal}
In this section, we consider what estimators are valid based on the idea
of Chapelle et al. and which estimator is `good' among valid estimators.  
To do so, let us calculate the bias correction factor $T^*(n,d)$ without
the orthonormal assumption (\ref{orthassumption}). 
As described before, the invariance of DEE under coordinate transformation
was obtained based on the inappropriate way. 
   \begin{theorem}\label{Tcalcnonorth}
    Let $\phi_1(x)\equiv 1$. Suppose that only the assumption A2 is
    satisfied in Theorem \ref{DEEderivation}.
Then $T^*(n,d)$
   is exactly calculated as
\begin{eqnarray}
 T^*(n,d)
    =\frac{1+(1/n)E_D[\trace(C\widehat{C}^{-1})]}{1-(d/n)}, \label{correction2}
\end{eqnarray}
   where $C=[C_{kk'}]$ is a $(d\times d)$ matrix with $C_{kk'}:=E_{x}[\phi_k(x)\phi_{k'}(x)]$.
  \end{theorem}
  See Section \ref{Tcalcnonorthproof} for its proof. The form of
  (\ref{correction2}) is invariant under coordinate transformation. It
  is natural because the definition of $T^*(n,d)$ is invariant under
  coordinate transformation due to the property of LSE. 
There remain two unknown quantities $C$ and $V:=E_D[\widehat{C}^{-1}]$
  in \eqnum{correction2}. Remarkably, both of them can be estimated
  using only the information about covariates $x$. 
  Let us define $D_0:=\{x_1,x_2,\cdots, x_n\}$ and $D_x:=D_0\cup D_U$. 
 Taking Theorem \ref{Tcalcnonorth} into account, it is natural to
  consider the following class of risk estimators. 
   \begin{definition}[a class of valid risk estimators]
We say that a risk estimator $\riskhat(d)$ is valid (in the sense of
    DEE) if there exists a consistent estimator\footnote{That is, $\widehat{H}(D_x)$ converges to the
    true value $CV$ in probability as $n$ and $n'$ go to the infinity.}
    $\widehat{H}(D_x)$ of $CV$ such that 
\begin{eqnarray}
 \riskhat(d)=\widehat{T}(n,d)\Remp(f_d(\cdot ;\hat{\alpha})),\quad
  \widehat{T}(n,d):=\frac{1+(1/n)\trace(\widehat{H}(D_x))}{1-(d/n)}. \nonumber
\end{eqnarray}
   \end{definition}
We can easily understand that the resultant form of DEE is valid under
  some regularity conditions because $\widetilde{C}$ and $\widehat{C}^{-1}$
  in (\ref{finaldee}) are statistically independent and are consistent
  estimators of $C$ and $V$ respectively.
  We imagine that Chapelle et al.\ knew the result of Theorem \ref{Tcalcnonorth} because they
  implied this result in Remark 3 of Section 2.2 of
  \citep{chapelleetal02}. Based on this fact, they seemingly recognized that the resultant
  form of DEE is valid. However, it is unclear that DEE is close to the
  optimal estimator in this class.
  In general, $V$ is more
  difficult to estimate than $C$ because $V$ is based on the inverse
  matrix of $\widehat{C}$. More concretely, $\widehat{C}^{-1}$ tends to fluctuate more
  largely than $\widehat{C}$ especially when $n$ is not large enough.
  Hence, spending more samples to estimate $V$ seems to be a reasonable
  strategy. However, DEE spends the most of $D_x$ (\thatis, $n'$ samples) to
  estimate $C$ and spends only $n$ samples to estimate $V$. Note that
  this strategy of DEE for estimation of $CV$ has no necessity since
  the corresponding part was derived inappropriately. Therefore, let us
  discuss what estimators are good in among valid risk estimators. We
  start from the following theorem.  
 \begin{theorem}\label{riskerror}
  Let $\riskhat$ be a valid risk estimator with $\widehat{H}(D_x)$.
Define
  $\riskhat^*(\hat{\alpha}):=T^*(n,d)\Remp(f_d{\cdot;\hat{\alpha}})$ with $T^*(n,d)$ in (\ref{correction2}). If
  $\widehat{H}(D_x)$ depends on only unlabeled 
  data \thatis, $\widehat{H}(D_x)=\widehat{H}(D_U)$, then 
\begin{eqnarray}
  \lefteqn{\hskip-8mm E_D\left[\left(\risk(d)-\riskhat(d)\right)^2\right]=
   E_D\Big[\left(\risk(d)-\riskhat^*(d)\right)^2\Big]}\nonumber\\
 &&\hskip-1cm+2\frac{{\rm bias}(\widehat{H})}{n-d}{\rm
  cov}(\riskhat^*(d),\Remp(f_d(\cdot;\hat{\alpha}(D))))+\frac{{\rm
  MSE}(\widehat{H})}{(n-d)^2}E_D\left[\Remp(f_d(\cdot;\hat{\alpha}(D)))^2\right].\label{riskerrordecom}
\end{eqnarray}
  Here, ${\rm cov}(X,Y)$ denotes usual covariance between $X$ and $Y$
  and 
  \begin{eqnarray*}
   {\rm bias}(\trace(\widehat{H})):=E[\trace(\widehat{H})-\trace(CV)],\ {\rm MSE}(\trace(\widehat{H})):=E\left[\left(\trace\left(\widehat{H}\right)-\trace\left(CV\right)\right)^2\right]\!.
  \end{eqnarray*}
 \end{theorem}
The first term of the right side of \eqnum{riskerrordecom} is independent of the
    choice of $\widehat{H}$.
    It expresses error of the
    ideally unbiased (but unknown) risk estimator $\riskhat^*(d)$.
    The
    second term is of order $O(1/n)$ while the third term is of order
   $O(1/n^2)$. Therefore, it is natural to use an unbiased estimator of $CV$ as $\widehat{H}$.
   If $\widehat{H}$ is unbiased,
   \begin{eqnarray}
 E_D\left[\!\left(\risk(d)\!-\!\riskhat(d)\right)^2\right]
&=&
E_D\!\Big[\!\!\left(\risk(d)\!-\!\riskhat^*(d)\right)^2\!\Big]\nonumber\\
&& +\frac{{\rm
   Var}\left(\widehat{H}\right)}{(n-d)^2}E_D\left[\Remp(f_d(\cdot;\hat{\alpha}(D)))^2\right], \label{unbiasedmse}
   \end{eqnarray}
   where $\mbox{Var}(\widehat{H})$ denotes a variance of
   $\trace(\widehat{H})$. That is, as long as $\widehat{H}$ is unbiased,
   $\widehat{H}$ having the smallest $\mbox{Var}(\widehat{H})$ gives the best
   performance.
This fact motivates us to consider the following estimator. 
Let us divide the unlabeled data set $D_U$ into two data sets
$D_U^1:=\{x'_1,x'_2,\cdots, x'_{n_1}\}$ and
$D_U^2:=\{x'_{n_1+1},x'_{n_1+2},\cdots, x'_{n_1+n_2}\}$ for estimating $C$
and $V$ respectively.
Furthermore, we divide $D_U^2$ into $B_2:=\lfloor n_2/n\rfloor$ subsets
such that the $b$-th subset is
\[
D_b:=\{x'_{(b-1)\cdot
n+1+n_1},x'_{(b-1)\cdot n+2+n_1},\cdots, x'_{b\cdot n+n_1}\}.
\]
As a result, each $D_b$ is an i.i.d.\  copy of $D_0$. Define
$\widehat{C}_b$ as an empirical correlation matrix of $\phi(x)=(\phi_1(x),\phi_2(x),\cdots,\phi_d(x))^T$ based on $D_b$.
Then, it is natural to estimate $V$ as
\[
\widehat{V}:=\frac{1}{B_2}\sum_{b=1}^{B_2}\widehat{C}^{-1}_{b}. \label{ECinv} 
\]
On the other hand, we can estimate $C$ using $D_U^1$
simply as
\begin{eqnarray*}
\widehat{C}_+&:=&\frac{1}{n_1}\left(\sum_{j=1}^{n_1}\phi(x'_j)\phi(x'_j)^T\right).
\end{eqnarray*}
Then, we simply make $\widehat{H}_1:=\widehat{C}_+\widehat{V}$. The resultant risk estimator is 
\begin{eqnarray}
\riskhat^{1}:=\frac{1+(1/n)\trace(\widehat{H}_1)}{1-(d/n)}\Remp(f_d(\cdot;\hat{\alpha}(D))). \label{modified}  
\end{eqnarray}
We refer to this modified version of DEE as mDEE1. There are other
possible variations depending on how to estimate $C$ and $V$ based on
unlabeled data. We prepare the three candidates shown in Table
\ref{mdeevariation}.  
\begin{table}[h]
 \caption{Variation of mDEE. This table expresses which data are used to
 estimate $C$ and $V$.}
\label{mdeevariation}
\begin{center}
\begin{tabular}{ccc}
\hline
 & data used for $\widehat{C}_+$ & data used for $\widehat{V}$\\
 mDEE1& $D_U^1$ & $D_U^2$\\ 
 mDEE2& $D_U^1$ & $D_U$\\ 
 mDEE3& $D_U$ & $D_U$\\ \hline
\end{tabular}
\end{center}
\end{table}
Both mDEE2 and mDEE3 construct $\widehat{C}_+,\widehat{V}$ and $\widehat{H}$
in the same way as mDEE1. We write $\widehat{H}$ used for mDEE1-3 as
$\widehat{H}_1,\widehat{H}_2$ and $\widehat{H}_3$. While mDEE1 has no overlapped samples between $\widehat{C}_+$ and
$\widehat{V}$, mDEE3 uses all unlabeled data to estimate both $C$ and
$V$. The other estimator mDEE2 is their intermediate. 
By checking some properties of these estimators, we have the following theorem.
  \begin{theorem}\label{biasvariancetheorem}
   Assume that both $n_1$ and $n_2$ can be divided by $n$. Let
   $B_1:=n_1/n$, $B_2:=n_2/n$ and $B:=n'/n$.
Let $\mu$ and $\nu$ be column vectors obtained by vectorizing $C$ and $V$
  respectively. Similarly, we vectorize $\widehat{C}_a$ and
   $\widehat{C}^{-1}_b$ as $\hat{\mu}_a$ and $\hat{\nu}_b$. We also define $\hat{\mu}$ and
   $\hat{\nu}$ as i.i.d.\ copies of $\hat{\mu}_a$ and $\hat{\nu}_b$. 
Then,
  \begin{eqnarray*}
 &&  {\rm bias}(\widehat{H}_1)=0,\ {\rm bias}(\widehat{H}_2)={\rm bias}(\widehat{H}_3)=\frac{d-\trace(CV)}{B},\\
&&   {\rm Var}(\trace(\widehat{H}_1))=
 \frac{\trace\left({\rm Var}(\hat{\mu}){\rm Var}(\hat{\nu})\right)}{B_1B_2}
    +\frac{\trace\left({\rm Var}(\hat{\nu})\mu\mu^T\right)}{B_2}+\frac{\trace\left({\rm
	       Var}(\hat{\mu})\nu\nu^T\right)}{B_1}.
  \end{eqnarray*}
   Furthermore, if we fix $B$ (or equivalently $n$ and $n'$), the variance of
   $\trace(\widehat{H}_1)$ is minimized by the ceiling or flooring of 
\begin{equation}
   B^*_1=
   \left\{
   \begin{array}{cc}
      \left(\frac{a_1-\sqrt{a_1a_2}}{a_1-a_2}\right)B&\quad {\rm if
       }\  a_1\neq a_2\\
    B/2 & \quad {\rm otherwise}\\
   \end{array}
   \right., \label{optimalb1}
\end{equation}
   where
\begin{eqnarray*}
 a_1\!\!:=\!\frac{\trace(\Var(\hat{\mu})\Var(\hat{\nu}))}{B}\!+\!\trace(\Var(\hat{\mu})\nu\nu^T),\ a_2\!:=\!\frac{\trace(\Var(\hat{\mu})\Var(\hat{\nu}))}{B}\!+\!\trace(\Var(\hat{\nu})\mu\mu^T).
\end{eqnarray*}
  \end{theorem}
  See Section \ref{biasvariancetheoremsec} for its proof. 
  The estimator mDEE1 seems to be the most reasonable estimator because its first
  order term $O(1/n)$ vanishes. Furthermore, $\mbox{MSE}(\widehat{H}_1)=\Var(\trace(\widehat{H}_1))$
  can be calculated explicitly as in Theorem
  \ref{biasvariancetheorem}. This is beneficial because the optimal
  balance of sample numbers used to estimate $C$ and $V$, \thatis, $B_1^*$, can be
  estimated as follows. By the above theorem, it suffices to estimate the
  quantities $a_1$ and $a_2$ in order to estimate $B_1^*$. Both
  quantities can be calculated if we know $\mu$, $\nu$,
  $\Var(\hat{\mu})$ and $\Var(\hat{\nu})$. We estimate them such as
  \begin{eqnarray*}
&&   \bar{\mu}=\frac{1}{B}\sum_{b=1}^B\hat{\mu}_b,\
 \bar{\nu}=\frac{1}{B}\sum_{b=1}^B\hat{\nu}_b,\ \\
&& \overline{\Var}(\hat{\mu})=\frac{1}{B-1}\sum_{b=1}^B(\hat{\mu}_b\hat{\mu}^T_b-\bar{\mu}\bar{\mu}^T),\
 \overline{\Var}(\hat{\nu})=\frac{1}{B-1}\sum_{b=1}^B(\hat{\nu}_b\hat{\nu}^T_b-\bar{\nu}\bar{\nu}^T).
  \end{eqnarray*}
Thus, we propose to choose the optimal $B_1$ as the rounded value of
\eqnum{optimalb1} with $a_1$ and $a_2$ calculated by using the above
empirical estimates. 

In contrast to the case of mDEE1, the estimators mDEE2 and mDEE3 admit the bias.
Therefore, we should discuss their performance through
(\ref{riskerrordecom}) instead (\ref{unbiasedmse}).
Then, we must care about MSE of $\widehat{H}$. 
Recall that MSE can be decomposed into bias and variance terms (for
example, see \citep{hastieetal01}), \thatis,   
\begin{eqnarray}
 MSE(\widehat{H})&=&E\left[\left(\trace(\widehat{H})-\trace(CV)\right)^2\right]\nonumber\\
&=&
E\left[\left(\trace\left(\widehat{H}\right)-E\left[\trace\left(\widehat{H}\right)\right]\right)^2\right]+\left(E\left[\trace\left(\widehat{H}\right)\right]-\trace(CV)\right)^2\nonumber\\
&=& \Var(\trace(\widehat{H})) + \mbox{bias}(\trace(\widehat{H}))^2.\nonumber 
\end{eqnarray}
The estimators mDEE2 and mDEE3 are developed to decrease the variance of
$\widehat{H}$ at cost of the bias increase. This is effective when the variance is much larger than bias. If the bias increases, the second
term of \eqnum{riskerrordecom} gets larger. When we can obtain unlabeled
data as many as we like, $\mbox{bias}(\trace(\widehat{H}))$ 
can be decreased to zero as $B\rightarrow \infty$ by Theorem
\ref{biasvariancetheorem}. MSE also decreases to zero as $B\rightarrow
\infty$ because of the consistency of $\widehat{H}$. Nevertheless,
if the number of unlabeled data is not enough, we have to
take care which is better mDEE1 or mDEE2-3. 

Many readers may think that the variance of $\trace(\widehat{H}_2)$
should be calculated to estimate the optimal $B_1$ for mDEE2.
Surely, it is possible. However, the resultant form is
excessively complicated and includes the third and the fourth cross moments. 
Hence, the resultant way to choose $B_1$ is also computationally
expensive and instable when $d$ gets large. Thus, we do not employ the
exact variance evaluation and survey the performance of mDEE2 by
numerical simulations.

Finally, we review again DEE. DEE does not satisfy
the assumption of Theorem \ref{riskerror} because $\widehat{H}$ of DEE utilizes $D_0$. When $\widehat{H}$ is allowed to depend on $D_0$, 
then we cannot obtain any clear result like Theorem \ref{riskerror}. 
Hence, we cannot compare DEE and mDEE through Theorem \ref{riskerror}.
However, DEE looks similar to mDEE1 when $B_1=B-1$ is selected. Since
$B_1$ of mDEE1 is optimized by the above way, mDEE1 does not necessarily
behave similarly to DEE. 
Actually, it will be turned out by numerical experiments that the
estimated $B_1^*$ for mDEE1 tends to be very small compared to $B$.
This fact indicates that the most unlabeled data should be used to
estimate $V$. As a result, DEE does not exploit available data efficiently. 
\section{Proofs of Theorems}
In this section, we provide proofs to all original theorems. 
\subsection{Proof of Theorem \ref{Tcalcnonorth}}\label{Tcalcnonorthproof}
  \begin{proof}
We do not need to trace the whole derivation of DEE. The result for
non-orthonormal cases is obtained by using \eqnum{predee} as follows.  
For convenience, let $\phi(x):=(\phi_1(x),\phi_2(x),\cdots,
\phi_d(x))^T$ for each $d$. Because the basis is not orthonormal,
$C=E_x[\phi(x)\phi(x)^T]$ is not an identity matrix. 
Using $C$, define 
\[
 \phi'(x):=C^{-1/2}\phi(x).
\]
Then, $\{\phi'_k(x)|k=1,2,\cdots, d\}$ comprises an orthonormal basis of
   $\mbox{Span}(\phi)$.
   Note that the LSE estimate of regression function does not change if
   we replace the original basis $\{\phi_k(x)\}$ with this orthonormal
   basis $\{\phi_k'(x)\}$. 
Therefore, using the basis $\{\phi_k'(x)\}$, we obtain the same result
as \eqnum{correction2} except the replacement of $\widehat{C}$ with
$\widehat{C}'=(1/n)(\Phi')^T\Phi'$, where $\phi'$ is a matrix whose
$(i,k)$ element is $\phi_k'(x_i)$.  Noting that $\Phi'=\Phi C^{-1/2}$,
we can rewrite $\widehat{C}'$ as
\[
 \widehat{C}'=\frac{1}{n}C^{-1/2}\Phi^T\Phi C^{-1/2}=C^{-1/2}\widehat{C} C^{-1/2}.
\]
Substituting this into \eqnum{correction2}, we obtain a new version of
\eqnum{correction2} as 
\[
 T^*(n,d)=\frac{1+(1/n)\trace(E_D[C^{1/2}\widehat{C}^{-1}C^{1/2}])}{1-(d/n)}=\frac{1+(1/n)\trace(CE_D[\widehat{C}^{-1}])}{1-(d/n)} 
\]
for non-orthonormal cases.
  \end{proof}
\subsection{Proof of Theorem \ref{riskerror}}\label{riskerrorproof}
  \begin{proof}
The left side of (\ref{riskerrordecom}) is calculated as  
\begin{eqnarray}
 E_D\left[\left(\risk(d)-\riskhat(d)\right)^2\right]
  &=&
   E_D\Big[\left(\risk(d)-\riskhat^*(d)\right)^2+\left(\riskhat^*(d)-\riskhat(d)\right)^2\nonumber\\
&&
 +2\left(\risk(d)-\riskhat^*(d)\right)\left(\riskhat^*(d)-\riskhat(d)\right)\Big].\label{im1}
\end{eqnarray}
   The last term is calculated as
\begin{eqnarray*}
 \lefteqn{E\left[\left(\risk(d)-\riskhat^*(d)\right)\left(\riskhat^*(d)-\riskhat(d)\right)\right]}\\
  &=&E\left[\left(\risk(d)-\riskhat^*(d)\right)\left(T^*(n,d)-\widehat{T}(n,d)\right)\Remp(f_d(\cdot;\hat{\alpha}(D)))\right]\\
 &=&E\left[\left(\risk(d)-\riskhat^*(d)\right)\Remp(f_d(\cdot;\hat{\alpha}(D)))\right]E\left[\left(T^*(n,d)-\widehat{T}(n,d)\right)\right]\\
  &=&\mbox{cov}(\riskhat^*(d),\Remp(f_d(\cdot;\hat{\alpha}(D))))E\left[\left(\widehat{T}(n,d)-T^*(n,d)\right)\right].
\end{eqnarray*}
   The second last equality is obtained since $\widehat{T}(n,d)$ is
   statistically independent of $D$ ($\widehat{T}(n,d)$ depends only on $D_U$). The second term of \eqnum{im1} is
   calculated as 
\begin{eqnarray*}
 E\Big[\left(\riskhat^*(d)-\riskhat(d)\right)^2\Big]&=&E\left[\left(T^*(n,d)-\widehat{T}(n,d)\right)^2\Remp(f_d(\cdot;\hat{\alpha}(D)))^2\right]\\
&=&E\left[\left(T^*(n,d)-\widehat{T}(n,d)\right)^2\right]E\left[\Remp(f_d(\cdot;\hat{\alpha}(D)))^2\right].
\end{eqnarray*}
The proof is completed by noting that 
   \begin{eqnarray*}
    \widehat{T}(n,d)-T^*(n,d)
     =
     \frac{1+\frac{1}{n}\trace\left(\widehat{H}\right)-\left(1+\frac{1}{n}\trace\left(CV\right)\right)}{1-(d/n)}
     = \frac{\trace\left(\widehat{H}\right)-\trace\left(CV\right)}{n-d}.
   \end{eqnarray*}
  \end{proof}
  \subsection{Proof of Theorem \ref{biasvariancetheorem}}\label{biasvariancetheoremsec}
   \begin{proof}
Let us partition the whole unlabeled data set $D_U$ into subsets
  consisting of $n$ samples. We write them as $D_1,D_2,\cdots, D_B$.
Then, we can write $D^1_U=\cup_{a=1}^{B_1}D_a$ and
  $D^2_U=\cup_{b=B_1+1}^{B}D_b$. As before, each empirical correlation matrix based on $D_b$ is denoted by
    $\widehat{C}_b$. The bias of $\widehat{H}_1$ trivially vanishes
    because of the statistical independence between $\widehat{C}_+$ and $\widehat{V}$.
As for mDEE2, it holds that 
  \begin{eqnarray*}
 \widehat{H}_2&=& \left(\frac{1}{B_1}\sum_{a=1}^{B_1}\widehat{C}_a\right)\left(\frac{1}{B}\sum_{b=1}^{B}\widehat{C}^{-1}_b\right)=\frac{1}{B_1B}\sum_{a=1}^{B_1}\sum_{b=1}^{B}\widehat{C}_a\widehat{C}^{-1}_b.
 \end{eqnarray*}
 Taking expectation, we have
  \begin{eqnarray*}
 E[\widehat{H}_2]&=&
  \frac{1}{B_1B}\sum_{a=1}^{B_1}\sum_{b=1}^{B}E[\widehat{C}_a\widehat{C}^{-1}_b]
  =\frac{1}{B_1B}\sum_{a=1}^{B_1}\sum_{b=1}^{B}\left(\delta_{ab}I_d+(1-\delta_{ab})CV\right)\\
   &=&\frac{1}{B_1B}\left(B_1I_d+(BB_1-B_1)CV\right)=\frac{1}{B}\left(I_d+(B-1)CV\right).
  \end{eqnarray*}
  Hence, the bias of $\trace(\widehat{H}_2)$ is
  \[
   E[\trace(\widehat{H}_2)-\trace(CV)]=\frac{d-\trace(CV)}{B}.
  \]
  This does not depend on $B_1$, so that $\trace(\widehat{H}_3)$ has the
  same bias. Next, we calculate the variance of
    $\trace(\widehat{H}_1)$.
Let $\mathcal{B}_1 = \{ 1,2,...,B_1 \} $
and
    $\mathcal{B}_2 = \{ B_1+1, B_1 +2 ,..., B \}$.
    Since $\mathcal{B}_1$ and $\mathcal{B}_2$ are disjoint, 
$
E \mbox{Tr}(\hat{C}_a \hat{C}^{-1}_b) = \mu^T \nu$ for any $a\in \mathcal{B}_1$
    and $b\in \mathcal{B}_2$. Hence, we have
\begin{align*}
\mbox{Var}(\mbox{Tr}( \hat{H}_1))
= 
&E\Bigl(
\frac{1}{B_1B_2}
\sum_{a \in \mathcal{B}_1}\sum_{b \in \mathcal{B}_2} 
\mbox{Tr}(\hat{C}_a \hat{C}^{-1}_b)
-
\mu^T \nu
\Bigr)^2  \\
= 
&E\Bigl(
\frac{1}{B_1B_2}
\sum_{a \in \mathcal{B}_1}\sum_{b \in \mathcal{B}_2} 
\hat{\mu}_a^T \hat{\nu}_b
-
\mu^T \nu
\Bigr)^2  \\
=&
\frac{1}{B_1^2B_2^2}
E\Bigl(
\sum_{a \in \mathcal{B}_1}\sum_{b \in \mathcal{B}_2} 
\Bigl(
\hat{\mu}_a^T \hat{\nu}_b
-
\mu^T \nu
\Bigr)
\Bigr)^2 \\ \label{last}
=&
\frac{1}{B_1^2B_2^2}
\sum_{a \in \mathcal{B}_1}\sum_{b \in \mathcal{B}_2} 
\sum_{c \in \mathcal{B}_1}\sum_{d \in \mathcal{B}_2} 
E
\Bigl(
\hat{\mu}_a^T \hat{\nu}_b
-
\mu^T \nu
\Bigr)
\Bigl(
\hat{\mu}_c^T \hat{\nu}_d
-
\mu^T \nu
\Bigr).
\end{align*}
We will make a case argument for the terms in the last summation.
If $c \neq a$ and $b \neq d$,
both factors are independent of each other. Hence, we have
\[
E
\Bigl(
\hat{\mu}_a^T \hat{\nu}_b
-
\mu^T \nu
\Bigr)
\Bigl(
\hat{\mu}_c^T \hat{\nu}_d
-
\mu^T \nu
\Bigr)
=0.
\]
If $c=a$ and $d \neq b$, we have
\begin{align*}
E
\Bigl(
\hat{\mu}_a^T \hat{\nu}_b
-
\mu^T \nu
\Bigr)
\Bigl(
\hat{\mu}_c^T \hat{\nu}_d
-
\mu^T \nu
\Bigr)
&=
E
\Bigl(
\hat{\mu}_a^T \hat{\nu}_b
-
\mu^T \nu
\Bigr)
\Bigl(
\hat{\mu}_a^T \hat{\nu}_d
-
\mu^T \nu
\Bigr)
\\
&=
E
\Bigl(
\hat{\mu}_a^T \nu
-
\mu^T \nu
\Bigr)
\Bigl(
\hat{\mu}_a^T \nu
-
\mu^T \nu
\Bigr)
\\
&=
E
(\hat{\mu}_a^T
-
\mu^T) \nu
(\hat{\mu}_a^T
-
\mu^T) \nu
\\
&=
\mbox{Tr}(\mbox{Var}(\hat{\mu})\nu \nu^T)
\end{align*}
Similarly, if $c \neq a$ and $d=b $, we have
\begin{align*}
  E
\Bigl(
\hat{\mu}_a^T \hat{\nu}_b
-
\mu^T \nu
\Bigr)
\Bigl(
\hat{\mu}_c^T \hat{\nu}_d
-
\mu^T \nu
\Bigr)
=
\mbox{Tr}(\mu \mu^T\mbox{Var}(\hat{\nu})).
\end{align*}
Finally, 
if 
$c=a$ and $b=d$,
 \begin{align*}
  \lefteqn{
E
\Bigl(
\hat{\mu}_a^T \hat{\nu}_b
-
\mu^T \nu
\Bigr)
\Bigl(
\hat{\mu}_c^T \hat{\nu}_d
-
\mu^T \nu
  \Bigr)}\\
&=
E
  (\hat{\mu}_a^T \hat{\nu}_b - \mu^T \nu )^2
 =
E
(
\hat{\mu}_a^T (\hat{\nu}_b - \nu )
+
(\hat{\mu}_a^T - \mu^T )\nu
)^2\\
&=
E
\Bigl(
(\hat{\mu}_a^T -\mu^T) (\hat{\nu}_b - \nu )
+\mu^T (\hat{\nu}_b - \nu )
+
(\hat{\mu}_a^T - \mu^T )\nu \bigr)^2.
 \end{align*}
Since the three terms in the last side are not correlated to each other,
we have
\begin{align*}
E
(\hat{\mu}_a^T \hat{\nu}_b - \mu^T \nu )^2
&=
E
\bigl(
(\hat{\mu}_a^T -\mu^T) (\hat{\nu}_b - \nu )
\bigr)^2
+
E
\bigl(
\mu^T (\hat{\nu}_b - \nu ) \bigr)^2
+
E
\bigl((\hat{\mu}_a^T - \mu^T )\nu \bigr)^2\\
&=
\mbox{Tr}(\mbox{Var}(\hat{\mu})\mbox{Var}(\hat{\nu}))
+
\mbox{Tr}(\mu \mu^T \mbox{Var}(\hat{\nu}))
+
\mbox{Tr}(\mbox{Var}(\hat{\mu})\nu \nu^T).
\end{align*}
Therefore, we have
\begin{align*}
\lefteqn{\mbox{Var}(\mbox{Tr}( \hat{H}_1))}\\
 &
 \!\!=\!\! 
\frac{1}{B_1^2B_2^2}
\Bigl(
B_1 B_2 \mbox{Tr}(\mbox{Var}(\hat{\mu})\mbox{Var}(\hat{\nu}))
+
B_1^2 B_2\mbox{Tr}(\mu \mu^T \mbox{Var}(\hat{\nu}))
+
B_1 B_2^2
\mbox{Tr}(\mbox{Var}(\hat{\mu})\nu \nu^T)
\Bigr)\\
 &
 \!\!=\!\! 
\frac{1}{B_1B_2}
\Bigl(
\mbox{Tr}(\mbox{Var}(\hat{\mu})\mbox{Var}(\hat{\nu}))
+
B_1\mbox{Tr}(\mu \mu^T \mbox{Var}(\hat{\nu}))
+
B_2
\mbox{Tr}(\mbox{Var}(\hat{\mu})\nu \nu^T)
\Bigr).
\end{align*}
  Finally, we minimize the variance in terms of $B_1$. Since $B$ is
    fixed, $B_2=B-B_1$. Using 
    $1/(B(B-B_1))=(1/B)(1/B+1/(B-B_1))$, $\mbox{Var}(\trace(\widehat{H}_1))$
    is rewritten as  
\begin{eqnarray*}
 \Var(\trace(\widehat{H}_1))&=&
  \frac{1}{B_1}\left(\frac{\trace\left(\mbox{Var}(\hat{\mu})\mbox{Var}(\hat{\nu})\right)}{B}
		+\trace\left(\mbox{Var}(\hat{\mu})\nu\nu^T\right)
			     \right)\\
&&  +
  \frac{1}{B-B_1}
  \left(
   \frac{\trace\left(\mbox{Var}(\hat{\mu})\mbox{Var}(\hat{\nu})\right)}{B}
   +\trace\left(\mbox{Var}(\hat{\nu})\mu\mu^T\right)
	 \right)\\
  &=& \frac{a_1}{B_1}+\frac{a_2}{B-B_1}.
\end{eqnarray*}
    By regarding $\Var(\trace(\widehat{H}_1))$ as a
    continuous function of $B_1$ and differentiating it, 
    \[
  \frac{d}{dB_1}\Var(\trace(\widehat{H}_1))=\frac{a_2}{(B-B_1)^2}-\frac{a_1}{B^2_1}.
  \]
  By setting this to zero, we obtain the second order equation of
  $B_1$. Its solution is
  \[
    \left(\frac{a_1\pm \sqrt{a_1a_2}}{a_1-a_2}\right)B.
  \]
It is easy to check that $(a_1+\sqrt{a_1a_2})/(a_1-a_2)\notin
  (0,1)$ while $(a_1-\sqrt{a_1a_2})/(a_1-a_2)\in (0,1)$. 
    Since $\Var(\trace(\widehat{H}_1))$ is convex in $B_1\in(0,B)$, the
    minimum is attained by $(a_1-\sqrt{a_1a_2})/(a_1-a_2)$. Therefore, the
    optimal integer $B_1$ is its ceiling or floor.
 \end{proof}
\section{Numerical Experiments}\label{sim}
By numerical experiments, we compare the performance of mDEE with the original DEE and ADJ and other existing methods.
We basically employ the same setting as \citet{chapelleetal02}. 
Define Fourier basis functions $\phi_k:\Re\rightarrow \Re$ as 
\begin{eqnarray*}
 \phi_1(x)= 1,\ \phi_{2p}(x)=\sqrt{2}\cos(px),\ \phi_{2p+1}(x)=\sqrt{2}\sin(px).
\end{eqnarray*}
The regression model $f_d:\Re^M\rightarrow \Re$ is defined by
\begin{equation}
f_d(x;\alpha)=\sum_{m=1}^M\sum_{k=1}^d  \alpha_k\phi_k(x_m), \label{regfuncmodel}
\end{equation}
where $x_m$ denotes the $m$-th component of $x$. Note that
$f_d(x;\alpha)$ cannot span $L^2(\rs{X})$ even with $d\rightarrow
\infty$ when $M>1$. 
For each $d=1,2,\cdots, \bar{d}$, we compute LSE\footnote{To avoid the singularity of
$\Phi^T\Phi$, we used Ridge estimator. However,
its regularization coefficient is set to $\lambda=10^{-9}$. Therefore, it
almost works like LSE.} $\hat{\alpha}(D)$ for model $\rs{M}_d$.  
Then, we calculate various risk estimators (model selection criteria)
for each $d$ and choose $\hat{d}$ minimizing it.
The performance of each risk estimator is measured by so-called {\it regret}
defined by the log ratio of risk to the best model: 
\begin{equation}
 \log\left(\frac{\riskhat(f_{\hat{d}}(x;\hat{\alpha}(D)))}{\min_{d}\riskhat(f_d(x;\hat{\alpha}(D)))}\right).
\end{equation}
Here, $\riskhat(f_d(\cdot;\hat{\alpha}(d)))$ expresses a test error,
$\riskhat(f_d(x;\hat{\alpha}(d))):=\frac{1}{\bar{n}}\sum_{i=1}^{\bar{n}}(y''_i-f_d(x''_i;\hat{\alpha}(d)))^2$,
where the test data $\{(x_i'',y_i'')\sbar i=1,2,\cdots, \bar{n}\}$ are
generated from the same distribution as the training data.
We compare mDEE1-3 with FPE \citep{akaike70}, cAIC
\citep{sugiura78} and cv (five-fold
cross-validation) 
in addition to DEE and ADJ. 
In calculation of mDEE1, $B_1$ (or $n_1$)
was chosen according to the way described in Section \ref{proposal}. We
also used the same $B_1$ for mDEE2. 

\subsection{Synthetic Data}\label{numsim}
First, we conduct the same experiments as that of \citet{chapelleetal02}
as follows. 
We prepare the following two true regression functions,
\begin{eqnarray}
\mbox{sinc function}\ \  f_1(x)=\sin(4x)/4x,\quad 
 \mbox{step function}\ \ f_2(x)= I(x> 0),\nonumber
\end{eqnarray}
where $I(\cdot)$ is an indicator function returning $1$ if its argument
is true and zero otherwise. The sinc function can be approximated well by
fewer terms in \eqnum{regfuncmodel} compared to the step function. 
The training data are generated according to the regression model in
\eqnum{regmodel} with the above regression functions. The noise $\xi_i$
is subject to $N(\xi_i;0,\sigma^2)$ which denotes the normal distribution
with mean $0$ and variance $\sigma^2$. We prepare $n=10,20,50$ training
samples and $n'=1500$ unlabeled data. 
Covariates $x_i$ are generated from $N(0,\bar{\sigma}^2)$
independently in contrast to \citep{chapelleetal02}. Note that the above
basis functions are not orthonormal with respect to $p(x)$ in this case.   
The model candidate number $\bar{d}$ was chosen as $\bar{d}=8$ for
$n=10$, $\bar{d}=15$ for $n=20$ and $\bar{d}=23$ for $n=50$.
The number of test data is set as $\bar{n}=1000$ in all simulations.
We conducted a series of experiments by varying the regression function
and the sample number $n$ summarized in Table \ref{simsetting}. In each
experiment, $\sigma^2$ is varied among $\{0.01,0.05,0.1,0.2,0.3,0.4\}$. 
The experiments were
repeated $1000$ times. The results are shown in Tables
\ref{table3}-\ref{table8}. These tables show the median and IQR
(InterQuartile Range) of regret of each method.
\begin{table}
\caption{This table summarizes which table shows the result of which experiment.}
\label{simsetting}
\begin{center}
 \begin{tabular}{ccccccc}
  \hline
  Table & 3 & 4& 5& 6& 7& 8\\
  regression function & sinc & sinc & sinc & step &step & step\\
  $n$ & $10$ & $20$ & $50$ & $10$ & $20$ &$50$\\
\hline
 \end{tabular}
\end{center}
\end{table}
 \begin{table}[h]
  \caption{Median (IQR) of regret of each model selection method when the true
  regression function is $f_1(x)$, $n=10$.}
  \label{table3}
  \begin{tabular}{cccccccc}
   \hline
   $\sigma^2$& 0.01 & 0.05 & 0.1 &0.2 & 0.3 & 0.4 \\
   \hline
FPE   & 1.030 (2.720) & 0.888 (3.070) & 0.880 (2.780) & 0.707 (2.690) & 0.697 (2.540) & 0.650 (2.910) \\ 
cAIC  & 1.250 (1.050) & 0.452 (0.456) & 0.281 (0.271) & 0.114 (0.160)  & 0.084 (0.127) & 0.058 (0.116) \\ 
ADJ   & 0.336 (0.949) & 0.190 (0.492) & 0.197 (0.379) & 0.116 (0.277) & 0.107 (0.296) & 0.103 (0.287) \\ 
cv    & 0.929 (1.160) & 0.427 (0.548) & 0.280 (0.373) & 0.146 (0.253) & 0.102 (0.214) & 0.085 (0.252) \\ 
DEE   & 0.497 (1.390) & 0.361 (0.718) & 0.298 (0.503) & 0.180 (0.358) & 0.150 (0.393) & 0.113 (0.375) \\ 
mDEE1 & 1.190 (1.050) & 0.483 (0.510) & 0.300 (0.340) & 0.144 (0.215) & 0.102 (0.171) & 0.077 (0.170) \\ 
mDEE2 & 1.190 (1.050) & 0.483 (0.511) & 0.300 (0.341) & 0.144 (0.215) & 0.102 (0.170) & 0.077 (0.170) \\ 
mDEE3 & 1.210 (1.050) & 0.479 (0.510) & 0.299 (0.330) & 0.140 (0.212) & 0.103 (0.169) & 0.076 (0.162) \\ 
   \hline
  \end{tabular}
 \end{table}
 \begin{table}[h]
  \caption{Median (IQR) of regret of each model selection method when the true
  regression function is $f_1(x)$, $n=20$.}
  \label{table4}
  \begin{tabular}{cccccccc}
   \hline
   $\sigma^2$& 0.01 & 0.05 & 0.1 &0.2 & 0.3 & 0.4 \\
   \hline
FPE   & 0.901 (3.630) & 0.626 (2.910) & 0.562 (2.760) & 0.469 (3.180) & 0.340 (2.320) & 0.325 (2.380) \\ 
cAIC  & 0.342 (0.620) & 0.204 (0.537) & 0.265 (0.393) & 0.236 (0.214) & 0.139 (0.146) & 0.112 (0.127) \\ 
ADJ   & 0.254 (0.660) & 0.164 (0.478) & 0.155 (0.324) & 0.151 (0.217) & 0.104 (0.178) & 0.099 (0.158) \\ 
cv    & 0.335 (0.698) & 0.220 (0.518) & 0.215 (0.390) & 0.199 (0.236) & 0.134 (0.185) & 0.121 (0.167) \\ 
DEE   & 0.179 (0.619) & 0.162 (0.433) & 0.148 (0.341) & 0.181 (0.249) & 0.152 (0.199) & 0.132 (0.175) \\ 
mDEE1 & 0.462 (0.468) & 0.179 (0.493) & 0.211 (0.375) & 0.205 (0.220) & 0.141 (0.160) & 0.124 (0.151) \\ 
mDEE2 & 0.462 (0.467) & 0.178 (0.494) & 0.213 (0.375) & 0.205 (0.220) & 0.140 (0.159) & 0.124 (0.149) \\ 
mDEE3 & 0.469 (0.427) & 0.181 (0.500) & 0.215 (0.379) & 0.210 (0.218) & 0.139 (0.152) & 0.123 (0.150) \\ 
   \hline
  \end{tabular}
 \end{table}
 \begin{table}[h]
  \caption{Median (IQR) of regret of each model selection method when the true
  regression function is $f_1(x)$, $n=50$.}
  \label{table5}
  \begin{tabular}{cccccccc}
   \hline
   $\sigma^2$& 0.01 & 0.05 & 0.1 &0.2 & 0.3 & 0.4 \\
   \hline
   
FPE   & 0.036 (0.247) & 0.068 (0.197) & 0.064 (0.207) & 0.055 (0.188) & 0.081 (0.210) & 0.077 (0.161) \\ 
cAIC  & 0.002 (0.033) & 0.041 (0.111) & 0.040 (0.071) & 0.037 (0.096) & 0.055 (0.133) & 0.068 (0.109) \\ 
ADJ   & 0.022 (0.120) & 0.060 (0.130) & 0.052 (0.124) & 0.058 (0.170) & 0.092 (0.160) & 0.080 (0.129) \\ 
cv    & 0.011 (0.075) & 0.051 (0.118) & 0.048 (0.095) & 0.043 (0.123) & 0.073 (0.162) & 0.071 (0.118) \\ 
DEE   & 0.009 (0.063) & 0.042 (0.112) & 0.047 (0.090) & 0.038 (0.102) & 0.053 (0.132) & 0.064 (0.114) \\ 
mDEE1 & 0.005 (0.041) & 0.028 (0.102) & 0.040 (0.076) & 0.035 (0.086) & 0.048 (0.120) & 0.061 (0.106) \\ 
mDEE2 & 0.005 (0.041) & 0.028 (0.101) & 0.040 (0.076) & 0.035 (0.088) & 0.048 (0.121) & 0.061 (0.106) \\ 
mDEE3 & 0.005 (0.042) & 0.027 (0.099) & 0.040 (0.076) & 0.036 (0.088) & 0.048 (0.120) & 0.062 (0.107) \\   \hline
  \end{tabular}
 \end{table}

 \begin{table}[h]
  \caption{Median (IQR) of regret of each model selection method when the true
  regression function is $f_2(x)$, $n=10$.}
  \label{table6}
  \begin{tabular}{cccccccc}
   \hline
$\sigma^2$& 0.01 & 0.05 & 0.1 &0.2 & 0.3 & 0.4 \\
   \hline
FPE   & 1.180 (3.650) & 0.936 (3.470) & 0.887 (3.390) & 0.737 (3.120) & 0.567 (2.580) & 0.581 (2.910) \\ 
cAIC  & 0.074 (0.563) & 0.145 (0.618) & 0.309 (0.551) & 0.255 (0.326) & 0.199 (0.259) & 0.150 (0.232) \\ 
ADJ   & 0.121 (0.471) & 0.142 (0.545) & 0.150 (0.503) & 0.159 (0.406) & 0.148 (0.326) & 0.145 (0.320) \\ 
cv    & 0.075 (0.447) & 0.096 (0.486) & 0.086 (0.469) & 0.152 (0.372) & 0.168 (0.318) & 0.156 (0.288) \\ 
DEE   & 0.146 (0.628) & 0.130 (0.522) & 0.118 (0.508) & 0.185 (0.427) & 0.190 (0.358) & 0.178 (0.333) \\ 
mDEE1 & 0.023 (0.215) & 0.013 (0.194) & 0.007 (0.257) & 0.052 (0.286) & 0.133 (0.282) & 0.124 (0.253) \\ 
mDEE2 & 0.023 (0.215) & 0.013 (0.197) & 0.008 (0.259) & 0.052 (0.286) & 0.133 (0.282) & 0.124 (0.253) \\ 
mDEE3 & 0.020 (0.198) & 0.012 (0.188) & 0.008 (0.259) & 0.052 (0.288) & 0.129 (0.282) & 0.124 (0.250) \\ 
   \hline
  \end{tabular}
 \end{table}
 \begin{table}[h]
  \caption{Median (IQR) of regret of each model selection method when the true
  regression function is $f_2(x)$, $n=20$.}
  \label{table7}
  \begin{tabular}{cccccccc}
   \hline
   $\sigma^2$& 0.01 & 0.05 & 0.1 &0.2 & 0.3 & 0.4 \\
   \hline
FPE   & 2.660 (5.540) & 1.390 (4.620) & 0.910 (3.990) & 0.587 (3.600) & 0.459 (3.050) & 0.402 (2.670) \\ 
cAIC  & 0.134 (0.186) & 0.052 (0.163) & 0.040 (0.150) & 0.033 (0.219) & 0.030 (0.263) & 0.064 (0.250) \\ 
ADJ   & 0.130 (0.220) & 0.091 (0.242) & 0.075 (0.217) & 0.089 (0.311) & 0.111 (0.302) & 0.122 (0.279) \\ 
cv    & 0.130 (0.184) & 0.079 (0.229) & 0.065 (0.217) & 0.068 (0.273) & 0.079 (0.287) & 0.105 (0.268) \\ 
DEE   & 0.181 (0.317) & 0.113 (0.315) & 0.078 (0.247) & 0.080 (0.279) & 0.078 (0.290) & 0.075 (0.262) \\ 
mDEE1 & 0.133 (0.177) & 0.073 (0.191) & 0.049 (0.164) & 0.040 (0.193) & 0.026 (0.205) & 0.038 (0.225) \\ 
mDEE2 & 0.133 (0.177) & 0.073 (0.191) & 0.049 (0.164) & 0.040 (0.193) & 0.025 (0.205) & 0.037 (0.226) \\ 
mDEE3 & 0.134 (0.173) & 0.073 (0.190) & 0.047 (0.155) & 0.039 (0.191) & 0.024 (0.203) & 0.030 (0.223) \\ 
   \hline
  \end{tabular}
 \end{table}
 \begin{table}[h]
  \caption{Median (IQR) of regret of each model selection method when the true
  regression function is $f_2(x)$, $n=50$.}
  \label{table8}
  \begin{tabular}{cccccccc}
   \hline
   $\sigma^2$& 0.01 & 0.05 & 0.1 &0.2 & 0.3 & 0.4 \\
   \hline
FPE   & 0.522 (1.550) & 0.273 (1.120) & 0.169 (0.571) & 0.087 (0.342) & 0.077 (0.274) & 0.072 (0.274) \\ 
cAIC  & 0.140 (0.217) & 0.095 (0.105) & 0.071 (0.077) & 0.053 (0.060) & 0.028 (0.075) & 0.013 (0.091) \\ 
ADJ   & 0.093 (0.125) & 0.076 (0.095) & 0.058 (0.089) & 0.049 (0.070) & 0.034 (0.106) & 0.039 (0.152) \\ 
cv    & 0.122 (0.153) & 0.092 (0.101) & 0.072 (0.088) & 0.051 (0.068) & 0.031 (0.093) & 0.024 (0.128) \\ 
DEE   & 0.150 (0.246) & 0.100 (0.131) & 0.080 (0.109) & 0.058 (0.081) & 0.038 (0.106) & 0.023 (0.137) \\ 
mDEE1 & 0.116 (0.156) & 0.084 (0.105) & 0.071 (0.085) & 0.055 (0.069) & 0.036 (0.087) & 0.020 (0.118) \\ 
mDEE2 & 0.116 (0.155) & 0.085 (0.105) & 0.071 (0.085) & 0.055 (0.070) & 0.036 (0.087) & 0.020 (0.118) \\ 
mDEE3 & 0.112 (0.146) & 0.084 (0.104) & 0.071 (0.084) & 0.055 (0.070) & 0.036 (0.087) & 0.019 (0.116) \\ 
   \hline
  \end{tabular}
 \end{table}
As for these synthetic data, the performance of mDEE1-mDEE3 are almost same. Hence, we do not discriminate them here. 
When the true regression function is easy to estimate (\thatis, $f_1$)
and the noise variance $\sigma^2$ is small enough, DEE performs
comparatively or a little bit dominated mDEE. Otherwise, all mDEE dominated
DEE. Especially, mDEE is more stable than DEE because IQR of mDEE is
usually smaller than that of DEE. 
Compared to ADJ, mDEE also performed better than ADJ except the case
where $\sigma^2$ is small and the true regression function is $f_1$.
This observation holds to some extent in comparison with other
methods. As a whole, mDEE is apt to be dominated by existing methods
when the regression function is easy to estimate and the noise variance
is almost equal to zero. In other cases, mDEE usually dominated other
methods. Finally, we remark that the estimated $B_1$ for mDEE1 
took its value usually around $1-20$. This fact indicates that
$V=E_D[\widehat{C}^{-1}]$ requires much more samples to estimate than $C$. 
\subsection{Real World Data}
We conducted the similar experiments on some real-world data sets from UCI
\citep{bachelichman13}, StatLib and DELVE benchmark databases as shown
in Table \ref{rwsetting}. We used again \eqnum{regfuncmodel} as the 
regression model. The number of model candidates $\bar{d}$ was determined by $\lceil
(n-1)/M \rceil$. We varied $n$ as $n=20,50$ in experiments. The total
number of unlabeled data $n'$ is described in Table \ref{rwsetting}. 
The test data number $\bar{n}$ was set to 'total data
number'$-(n+n')$ in each experiment. 
\begin{table}[h]
\caption{Properties of real world data sets and experimental setting.}
\label{rwsetting}
\begin{center}
 \begin{tabular}{ccccccc}
\hline
  data set & $\mbox{dim}(x)$ & total data number &$n'$& source\\
  concrete compressive strength (concrete) &$8$&$1030$ & $800$&UCI \\
  NO2 (NO2)&  $7$& $500$ & $350$&StatLib \\
  bank 8nm (bank) & $8$& $8192$&$1300$&DELVE \\
  pumadyn family 8nm (puma) & 8&8192&  1300&DELVE \\
  heating energy efficiency (eeheat) & 8&768 & 550&UCI \\
  cooling energy efficiency (eecool) &8&768 &  550&UCI \\
  abalone & 7 &4177 & 1300 & UCI \\
\hline
 \end{tabular}
\end{center}
\end{table}
\begin{figure}[h]
 \begin{minipage}{.485\linewidth}
  \subfigure[data: concrete, $n=20$]{\includegraphics[width=\linewidth,height=5cm]{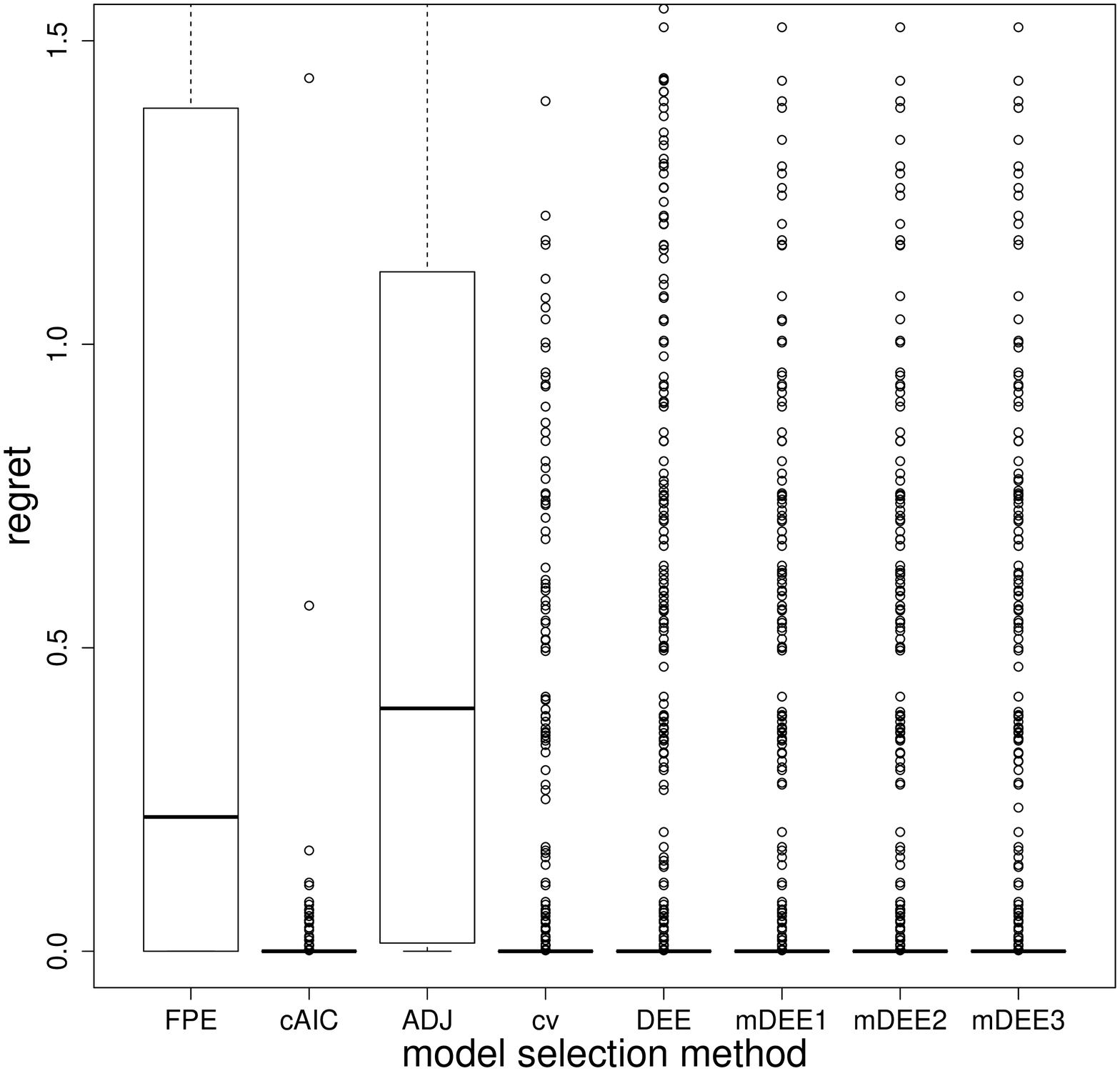}}
 \end{minipage}
 \begin{minipage}{.485\linewidth}
  \subfigure[data: concrete, $n=50$]{\includegraphics[width=\linewidth,height=5cm]{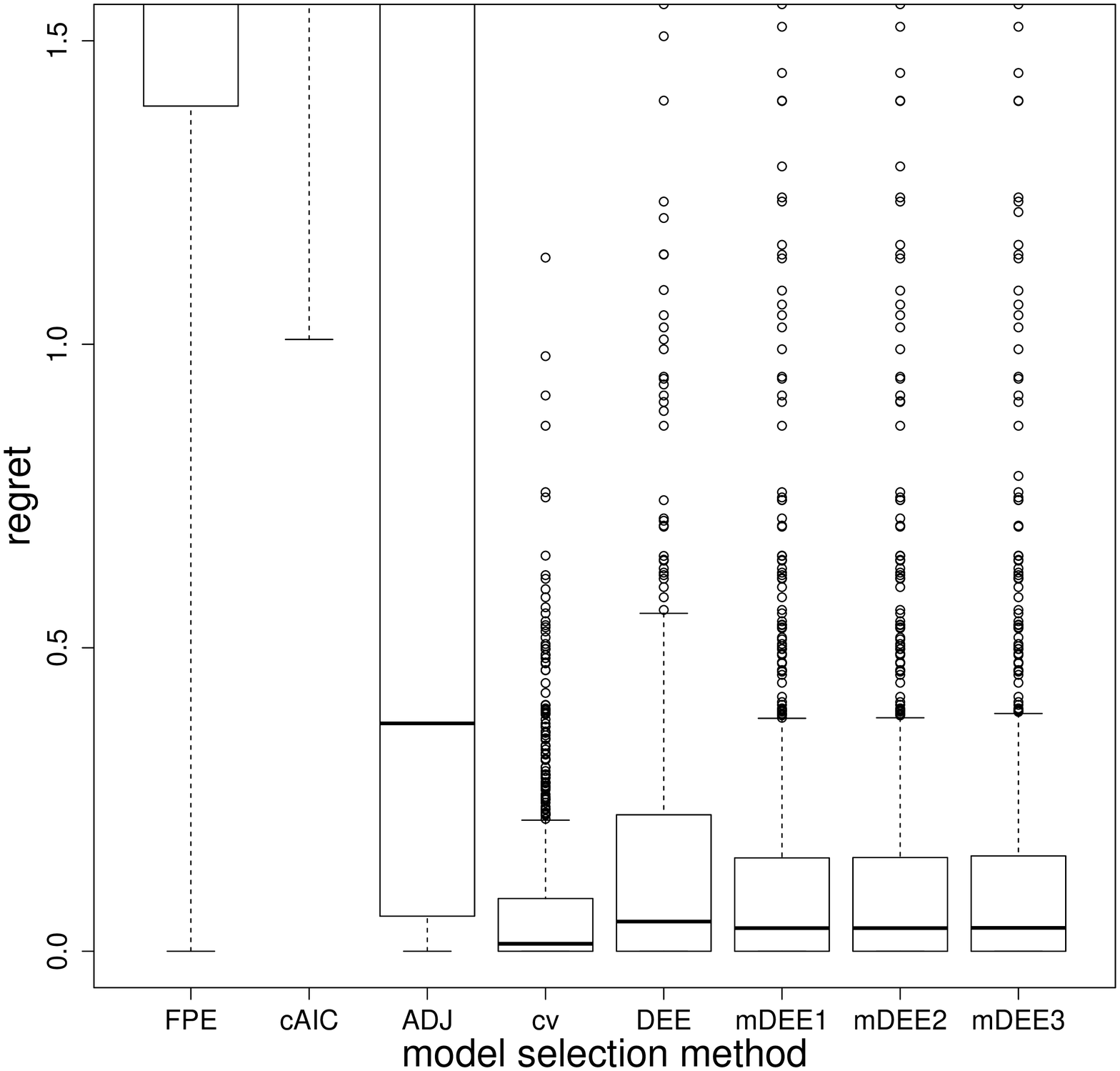}}
 \end{minipage}
 \begin{minipage}{.485\linewidth}
  \subfigure[data: NO2, $n=20$]{\includegraphics[width=\linewidth,height=5cm]{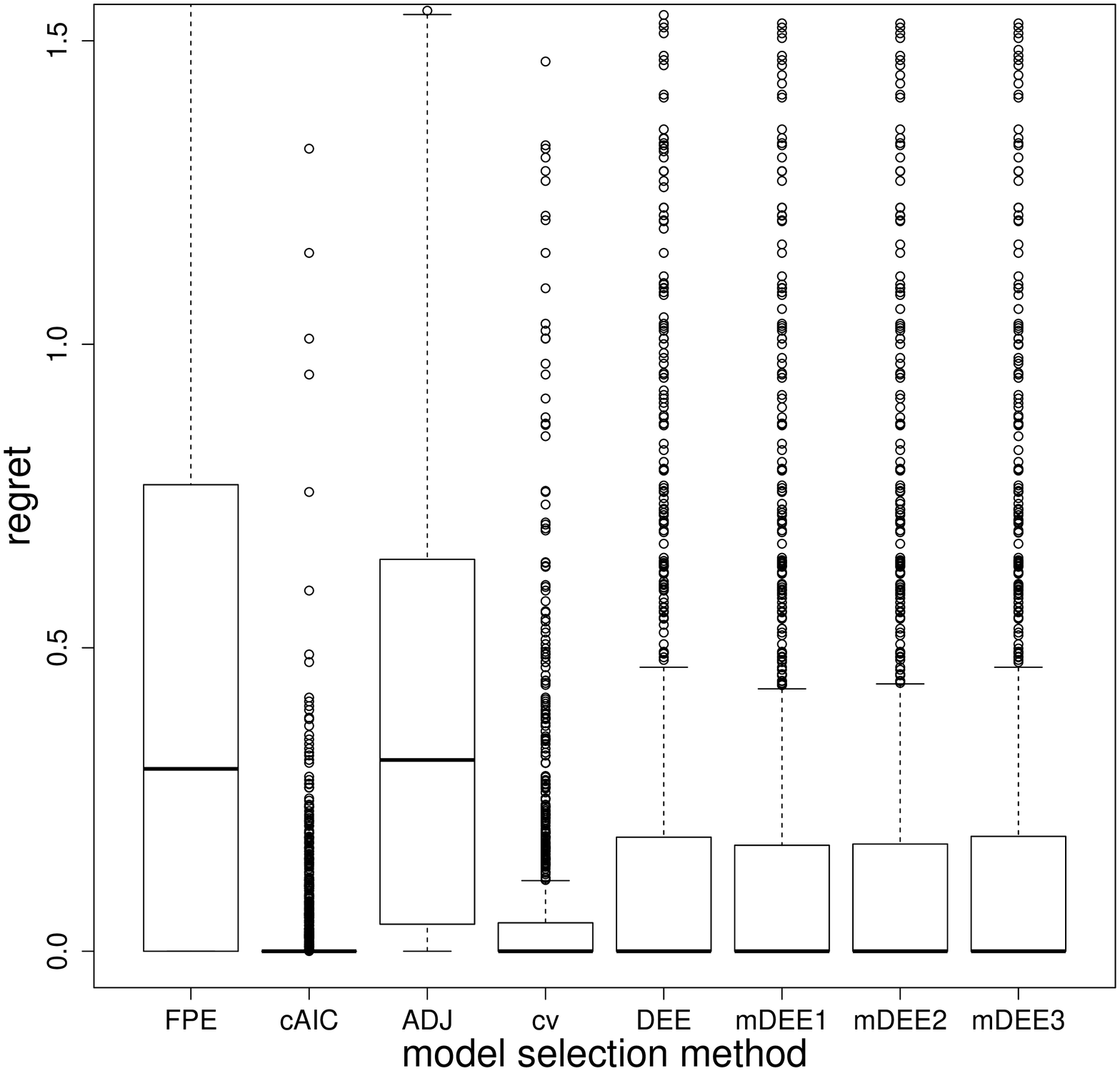}}
 \end{minipage}
 \begin{minipage}{.485\linewidth}
  \subfigure[data: NO2, $n=50$]{\includegraphics[width=\linewidth,height=5cm]{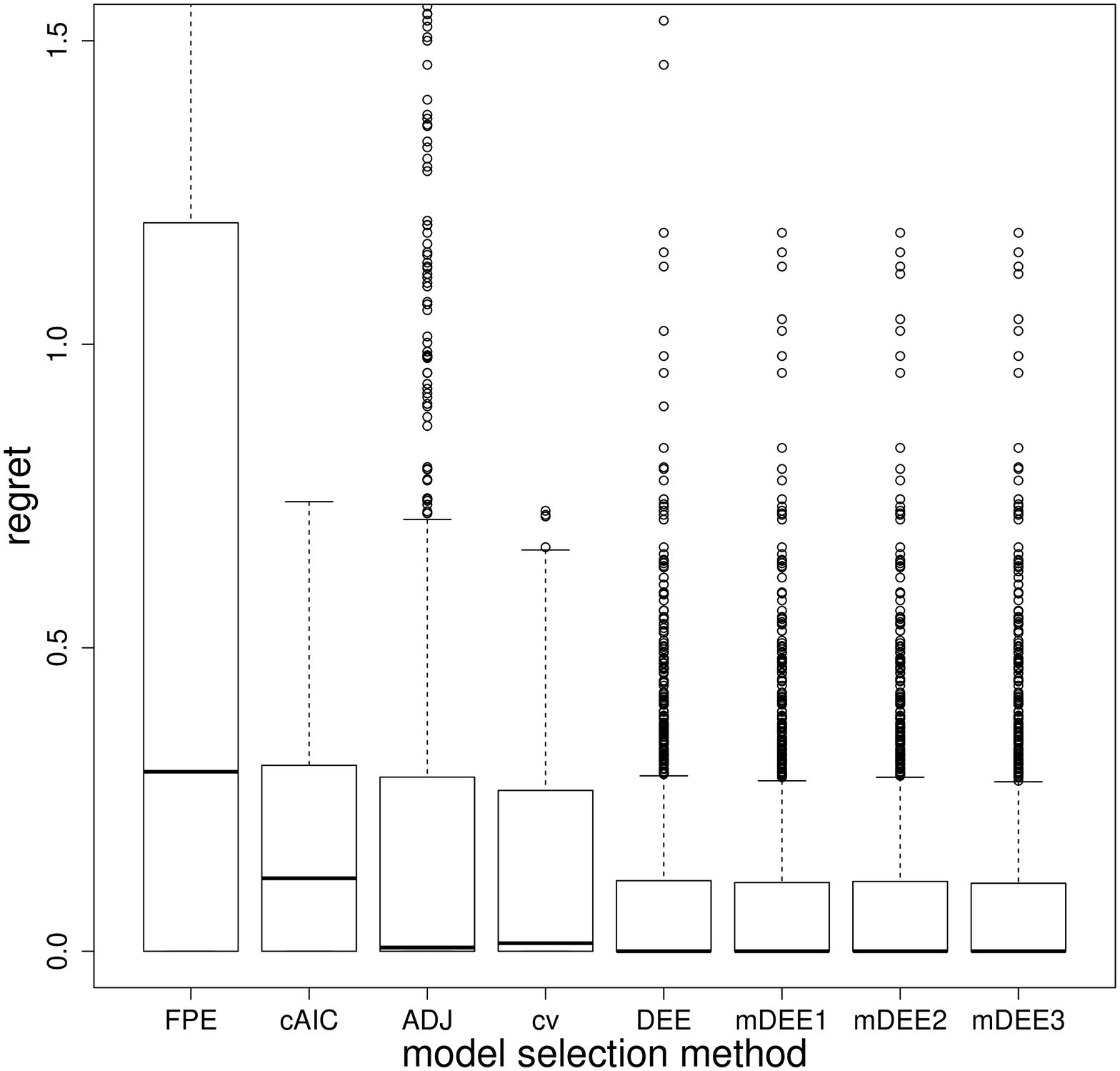}}
 \end{minipage}
 \begin{minipage}{.485\linewidth}
  \subfigure[data: bank, $n=20$]{\includegraphics[width=\linewidth,height=5cm]{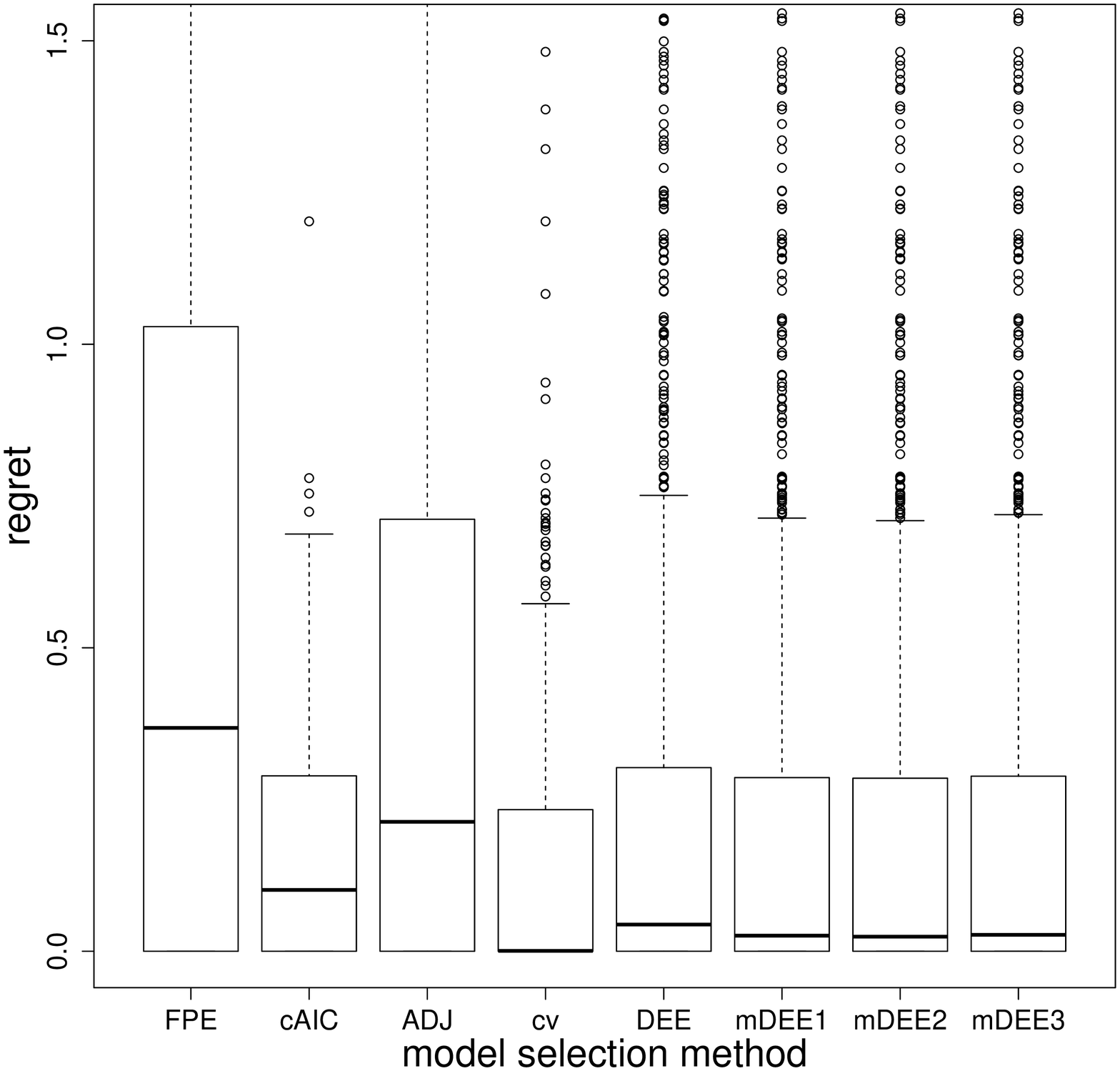}}
 \end{minipage}
 \begin{minipage}{.485\linewidth}
  \subfigure[data: bank, $n=50$]{\includegraphics[width=\linewidth,height=5cm]{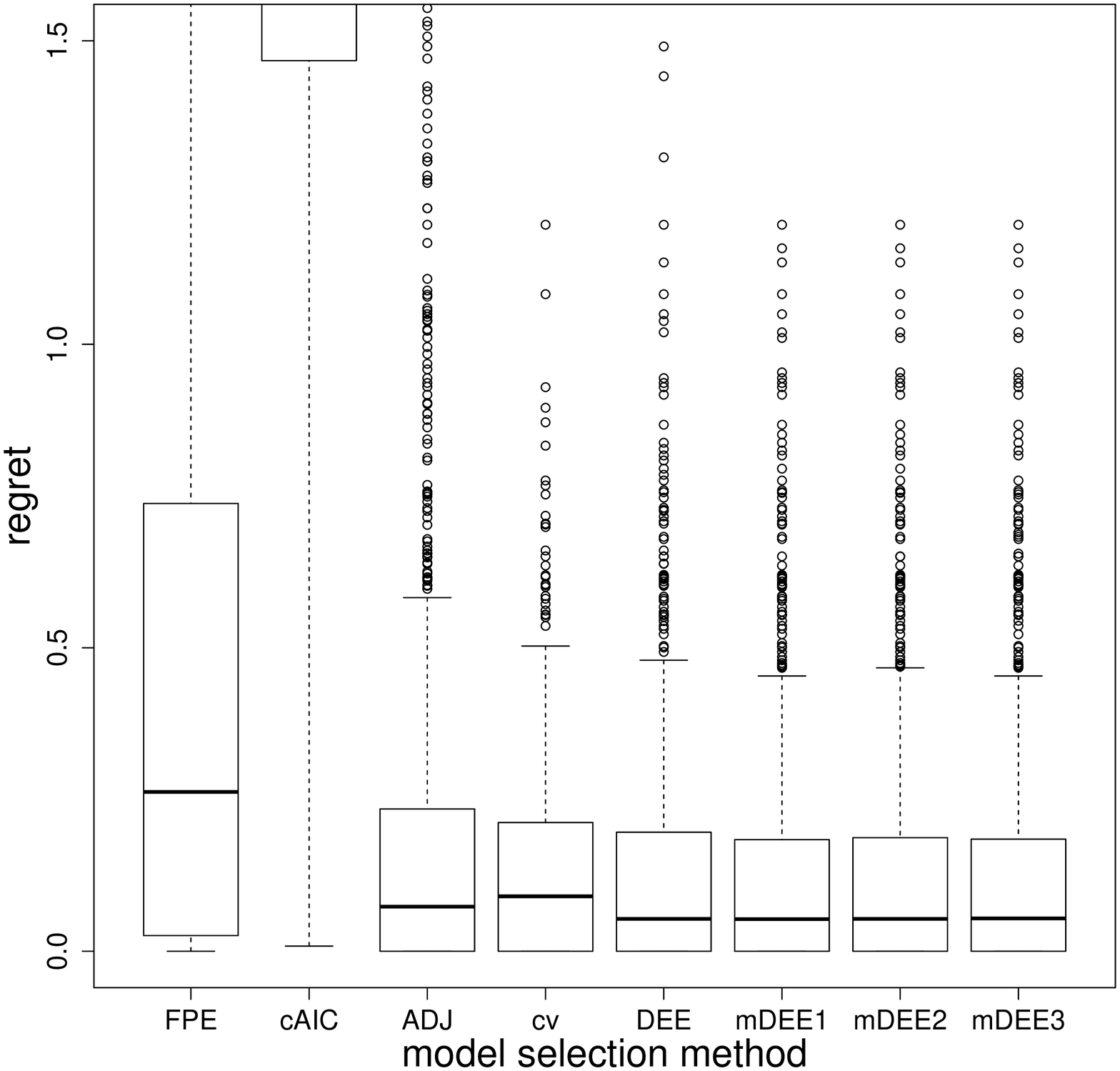}}
 \end{minipage}
 \begin{minipage}{.485\linewidth}
  \subfigure[data: puma, $n=20$]{\includegraphics[width=\linewidth,height=5cm]{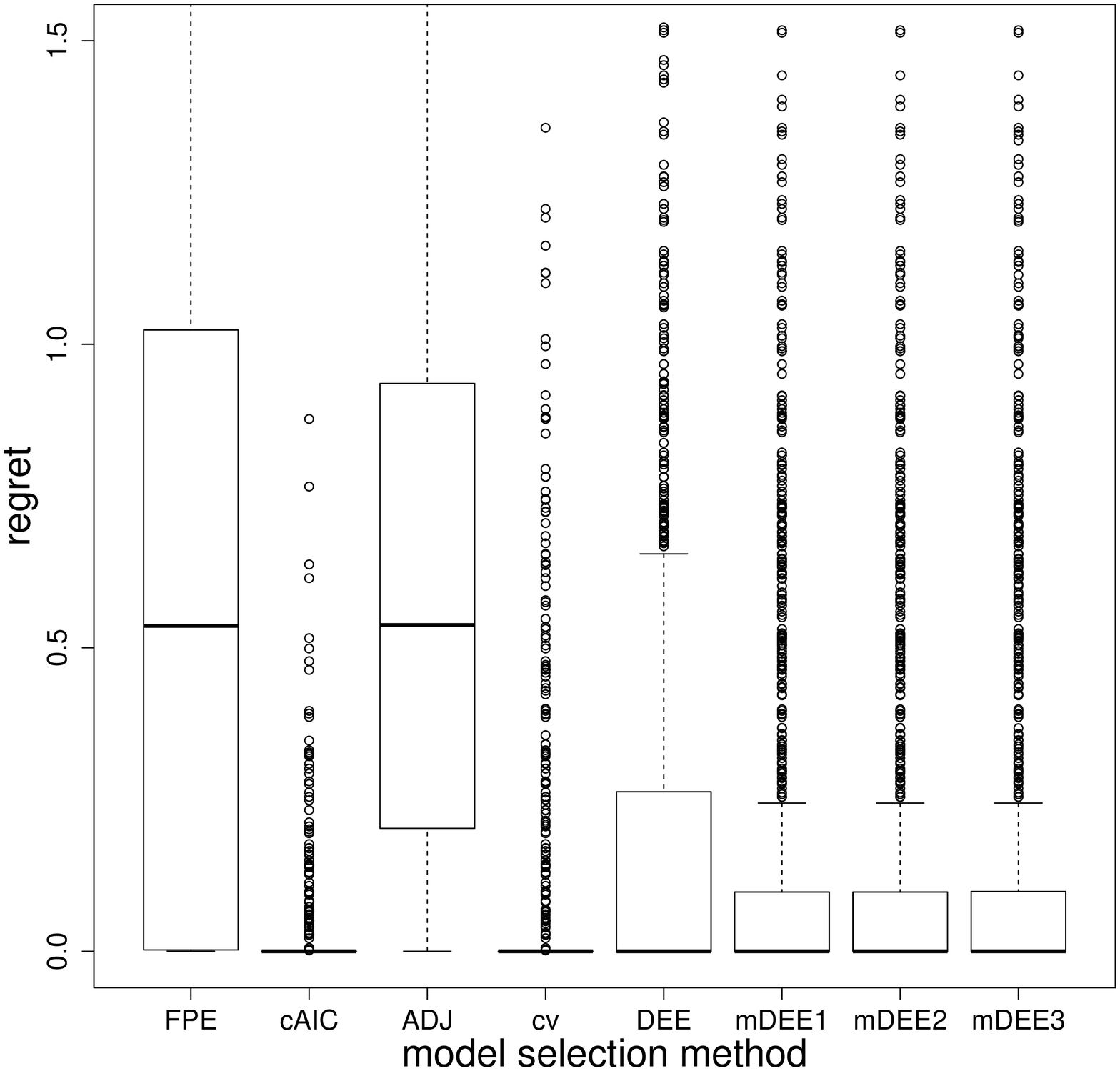}}
 \end{minipage}
 \begin{minipage}{.485\linewidth}
  \subfigure[data: puma, $n=50$]{\includegraphics[width=\linewidth,height=5cm]{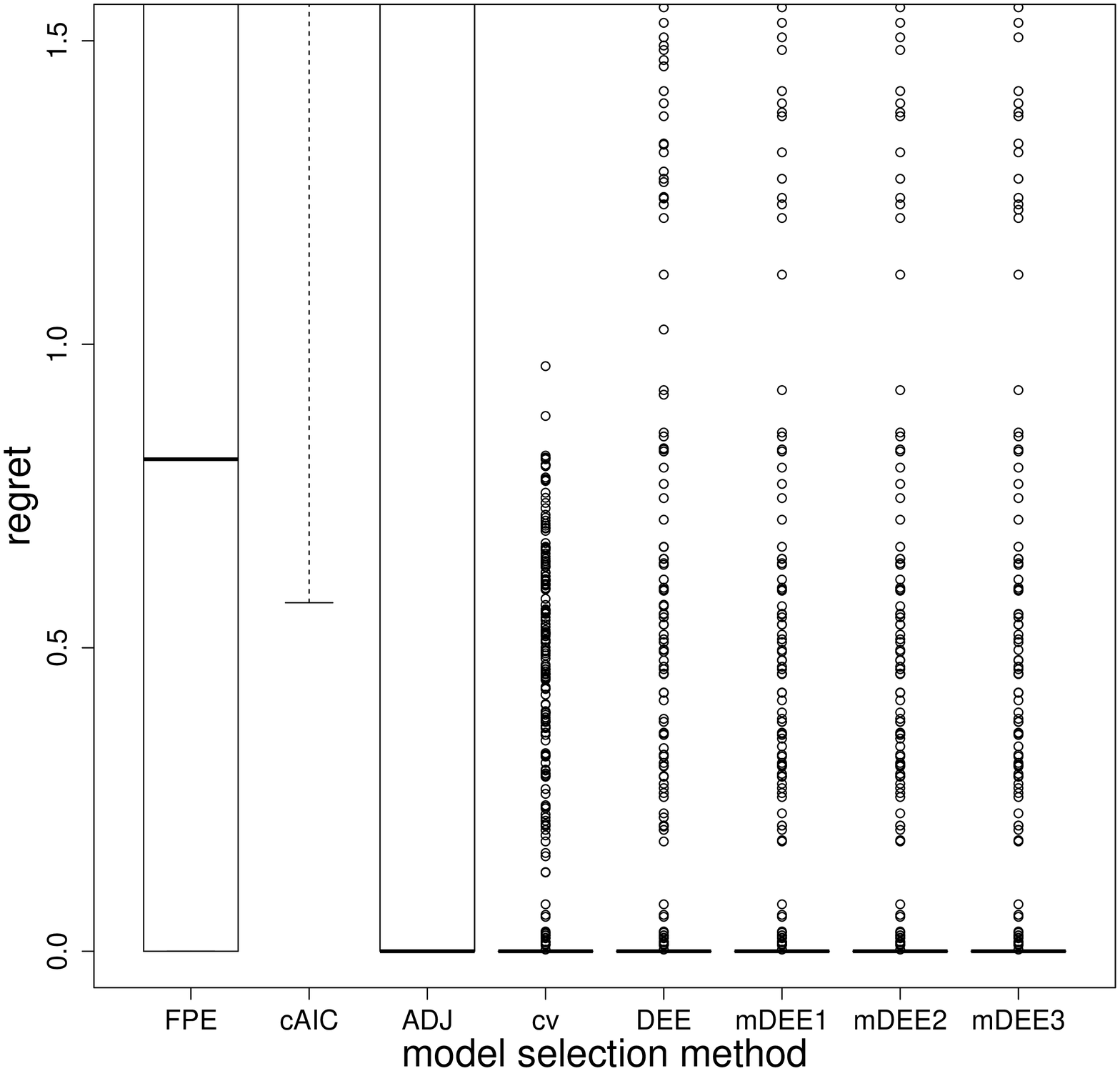}}
 \end{minipage}
\caption{Boxplot of regret for real-world data sets.}
\label{fig13}
\end{figure}

\begin{figure}[h]
 \begin{minipage}{.485\linewidth}
  \subfigure[data: eeheat, $n=20$]{\includegraphics[width=\linewidth,height=5cm]{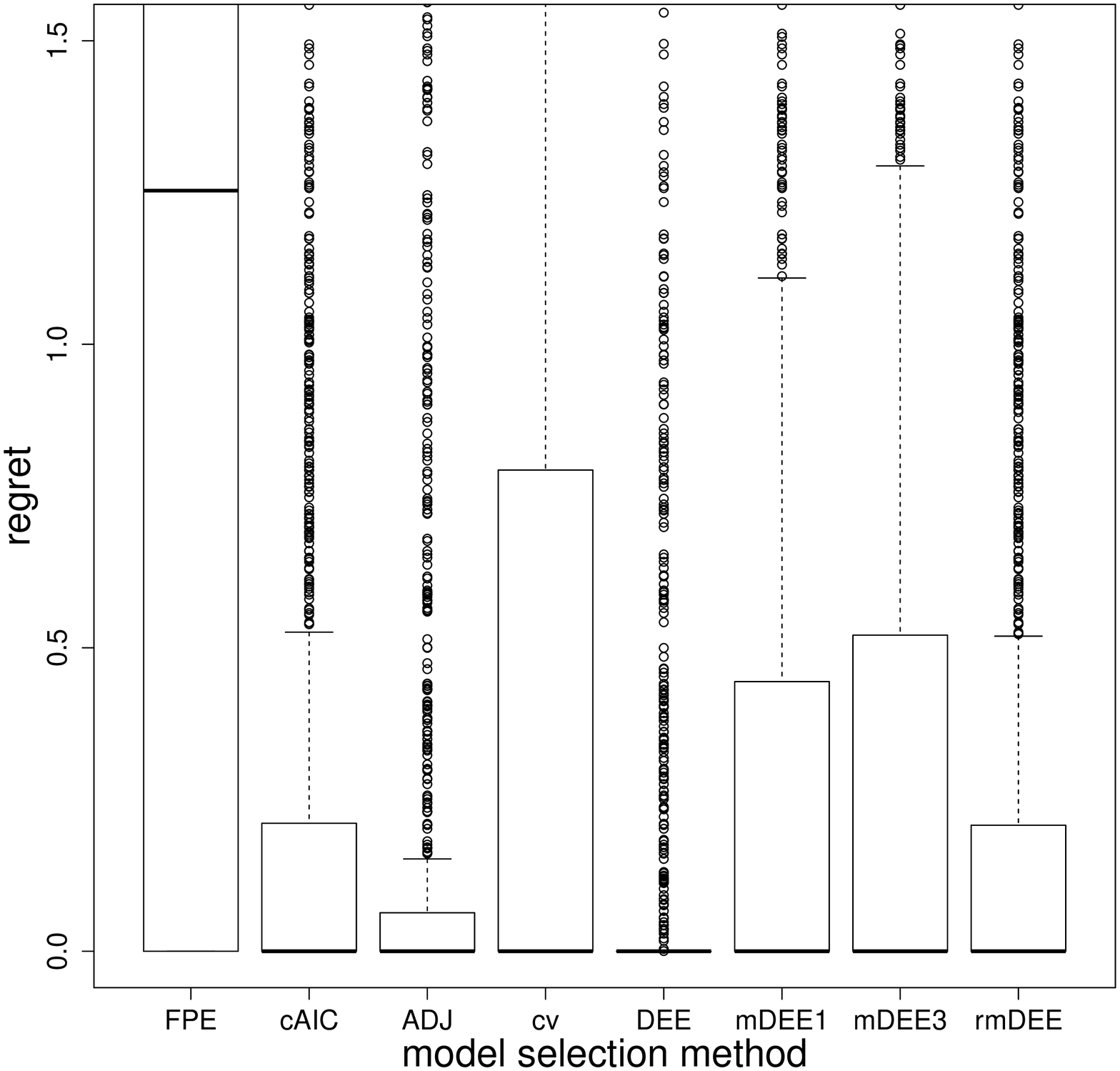}}
 \end{minipage}
 \begin{minipage}{.485\linewidth}
  \subfigure[data: eeheat, $n=50$]{\includegraphics[width=\linewidth,height=5cm]{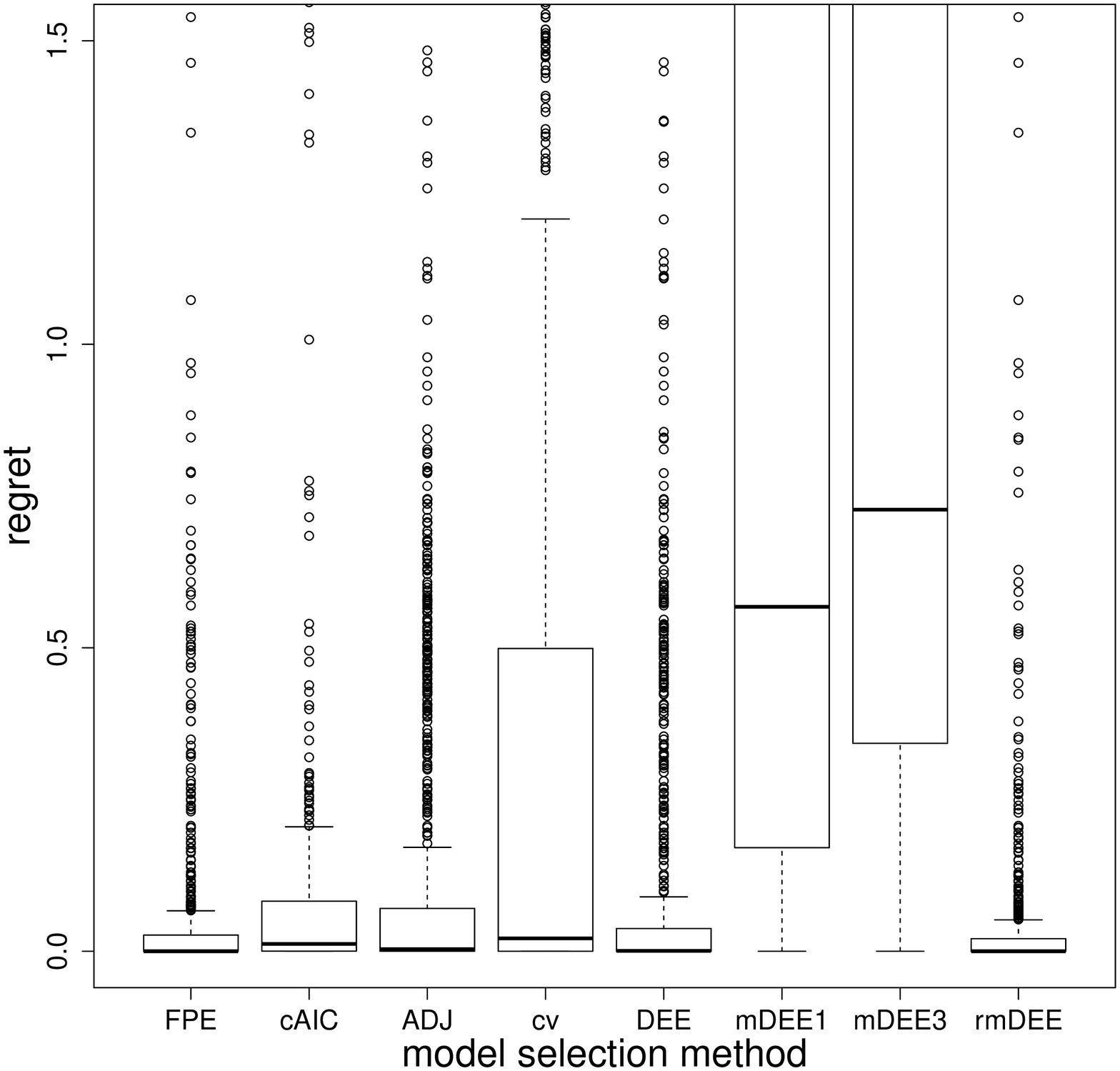}}
 \end{minipage}
 \begin{minipage}{.485\linewidth}
  \subfigure[data: eecool, $n=20$]{\includegraphics[width=\linewidth,height=5cm]{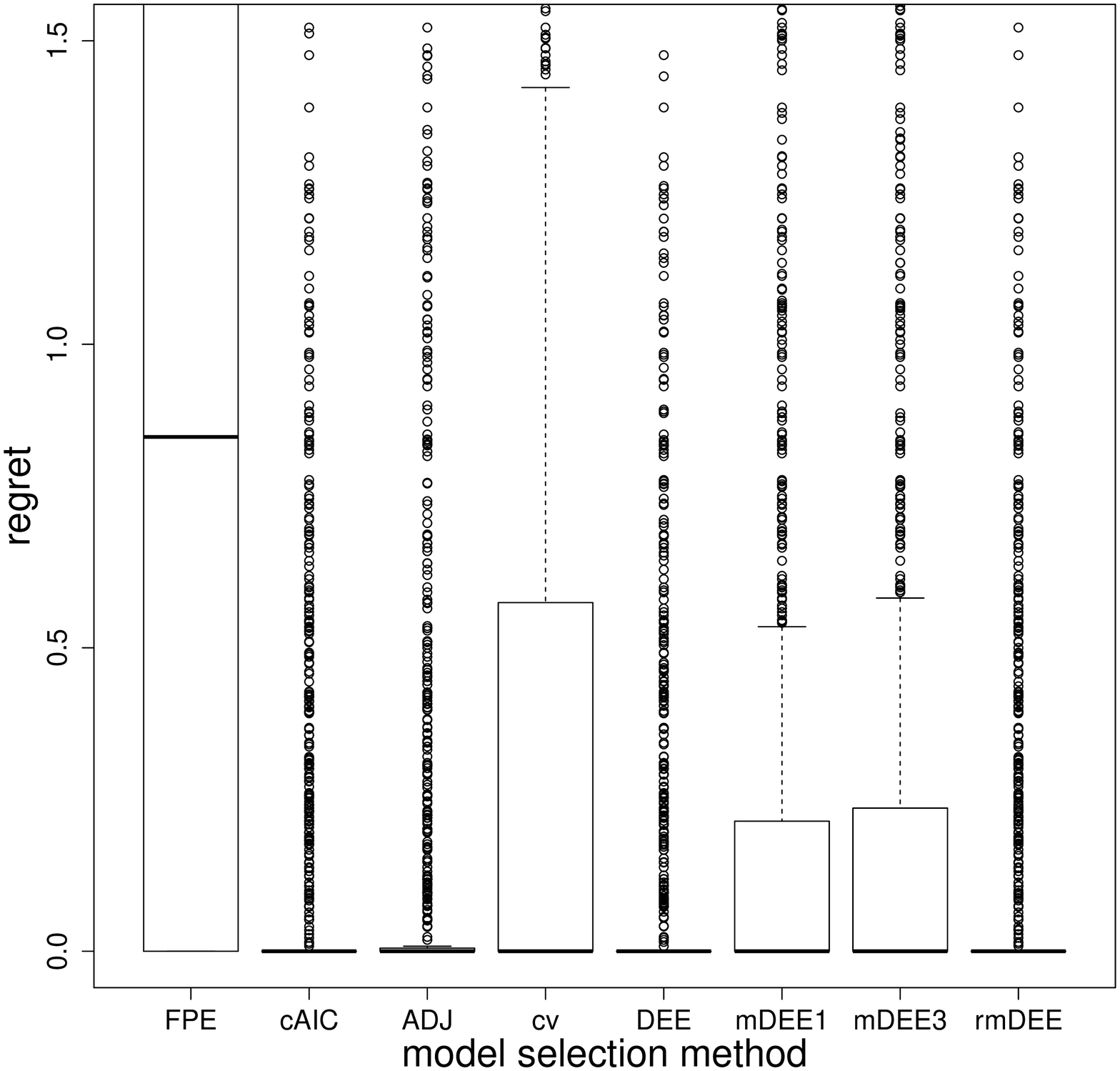}}
 \end{minipage}
 \begin{minipage}{.485\linewidth}
  \subfigure[data: eecool, $n=50$]{\includegraphics[width=\linewidth,height=5cm]{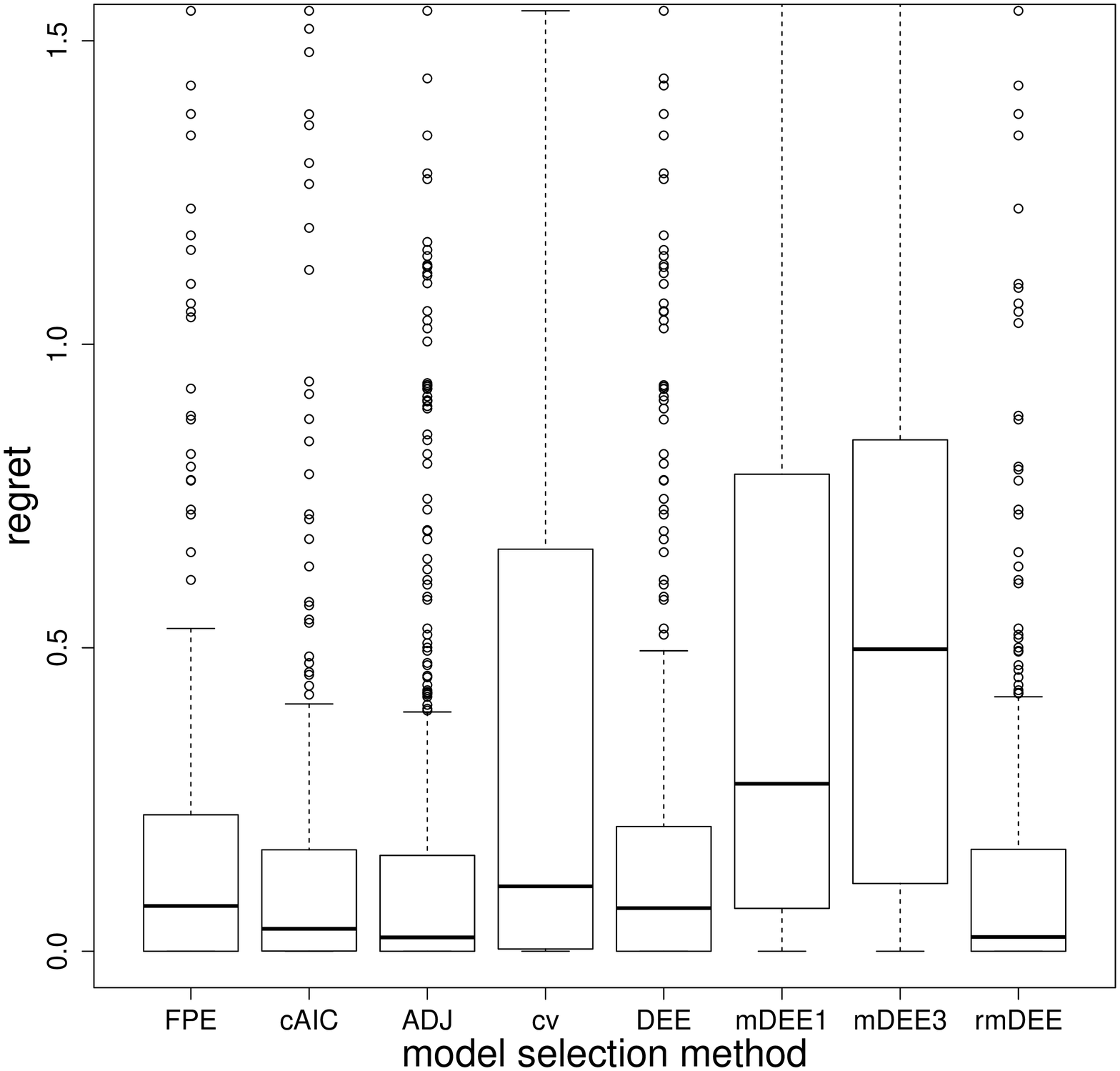}}
 \end{minipage}
 \begin{minipage}{.485\linewidth}
  \subfigure[data: abalone, $n=20$]{\includegraphics[width=\linewidth,height=5cm]{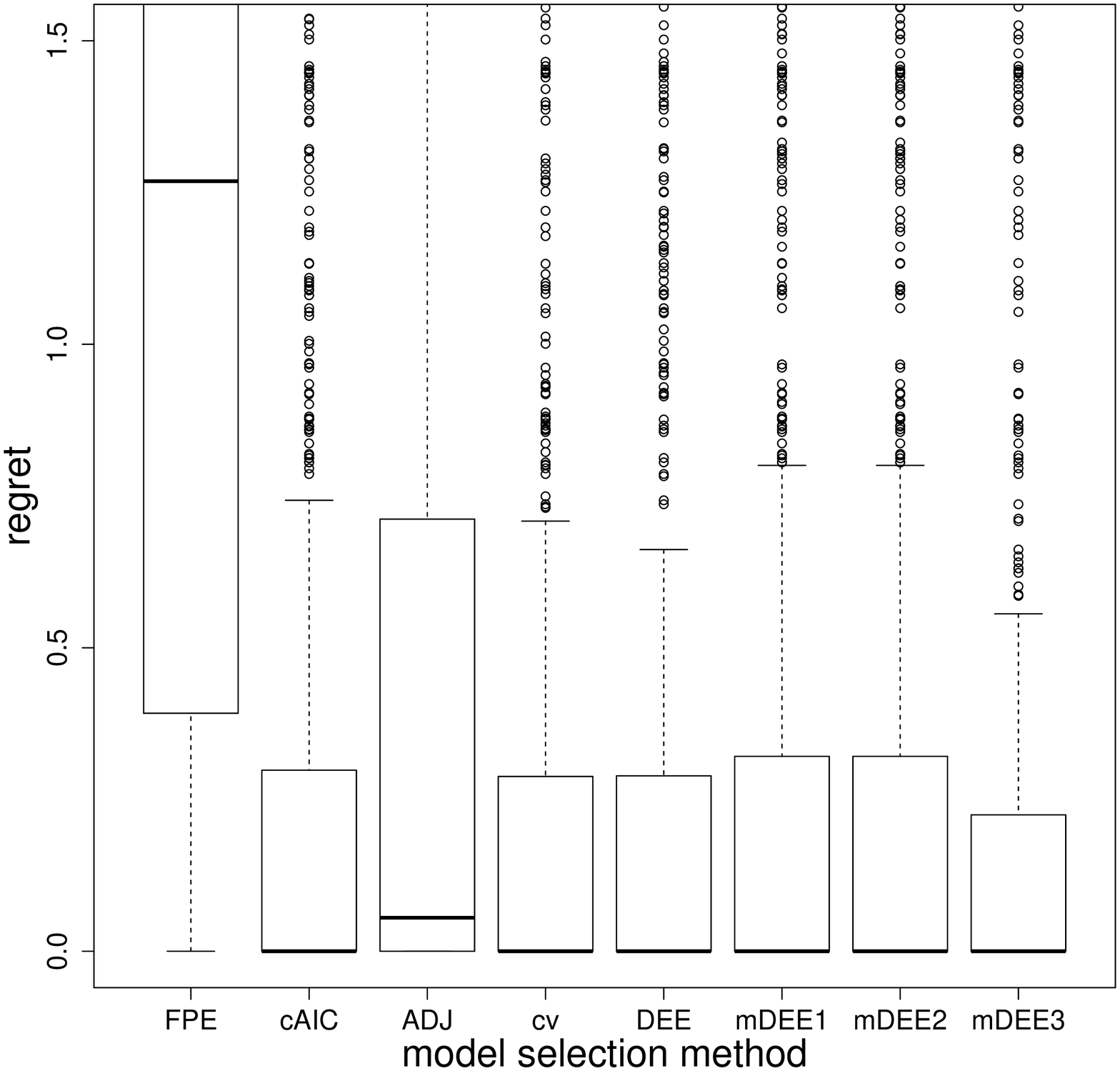}}
 \end{minipage}
 \begin{minipage}{.485\linewidth}
  \subfigure[data: abalone, $n=50$]{\includegraphics[width=\linewidth,height=5cm]{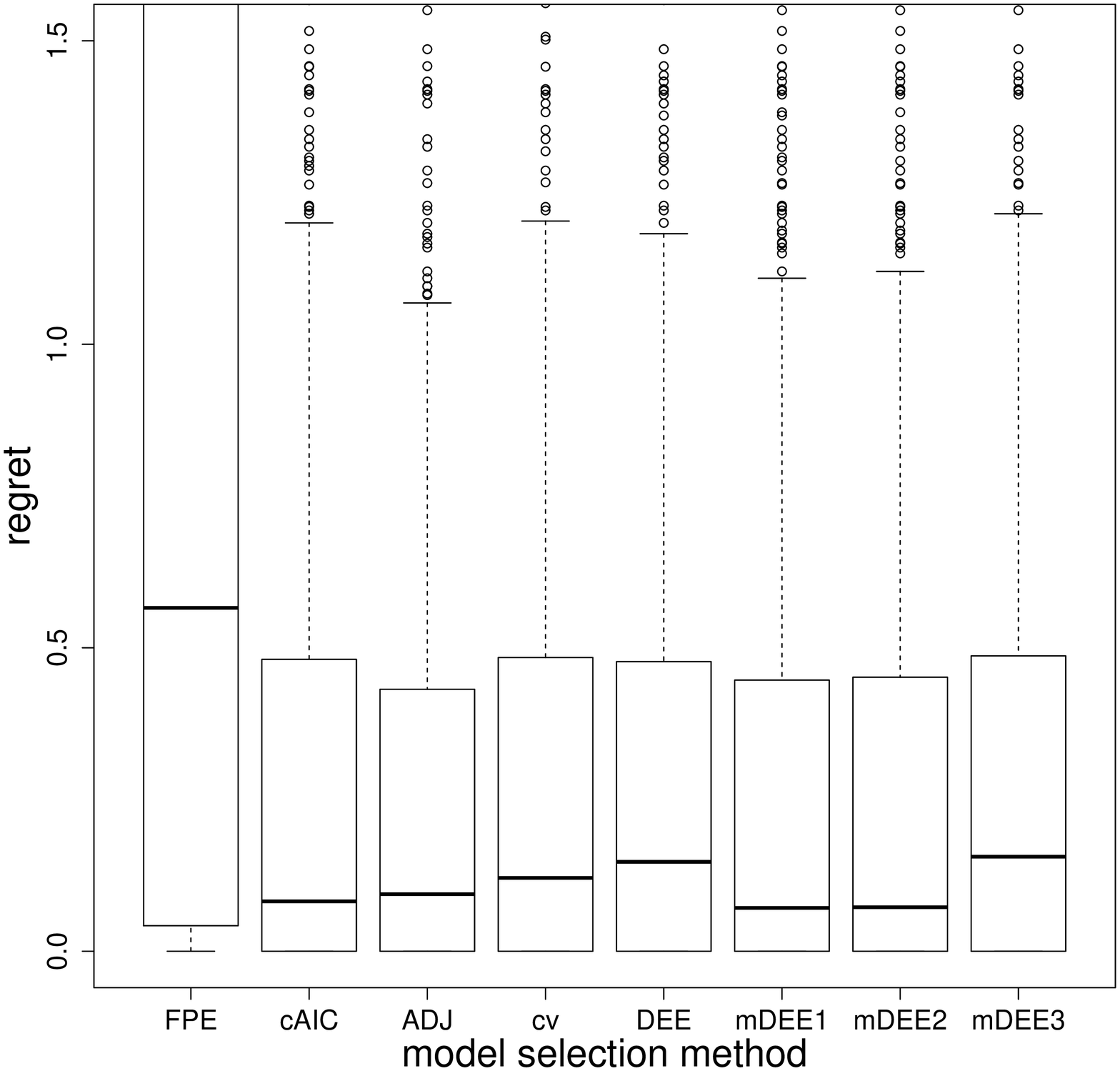}}
 \end{minipage}
\caption{Boxplot of regret for real-world data sets.}
\label{fig14}
\end{figure}
\begin{figure}[h]
 \begin{minipage}{.48\linewidth}
  \includegraphics[width=\linewidth]{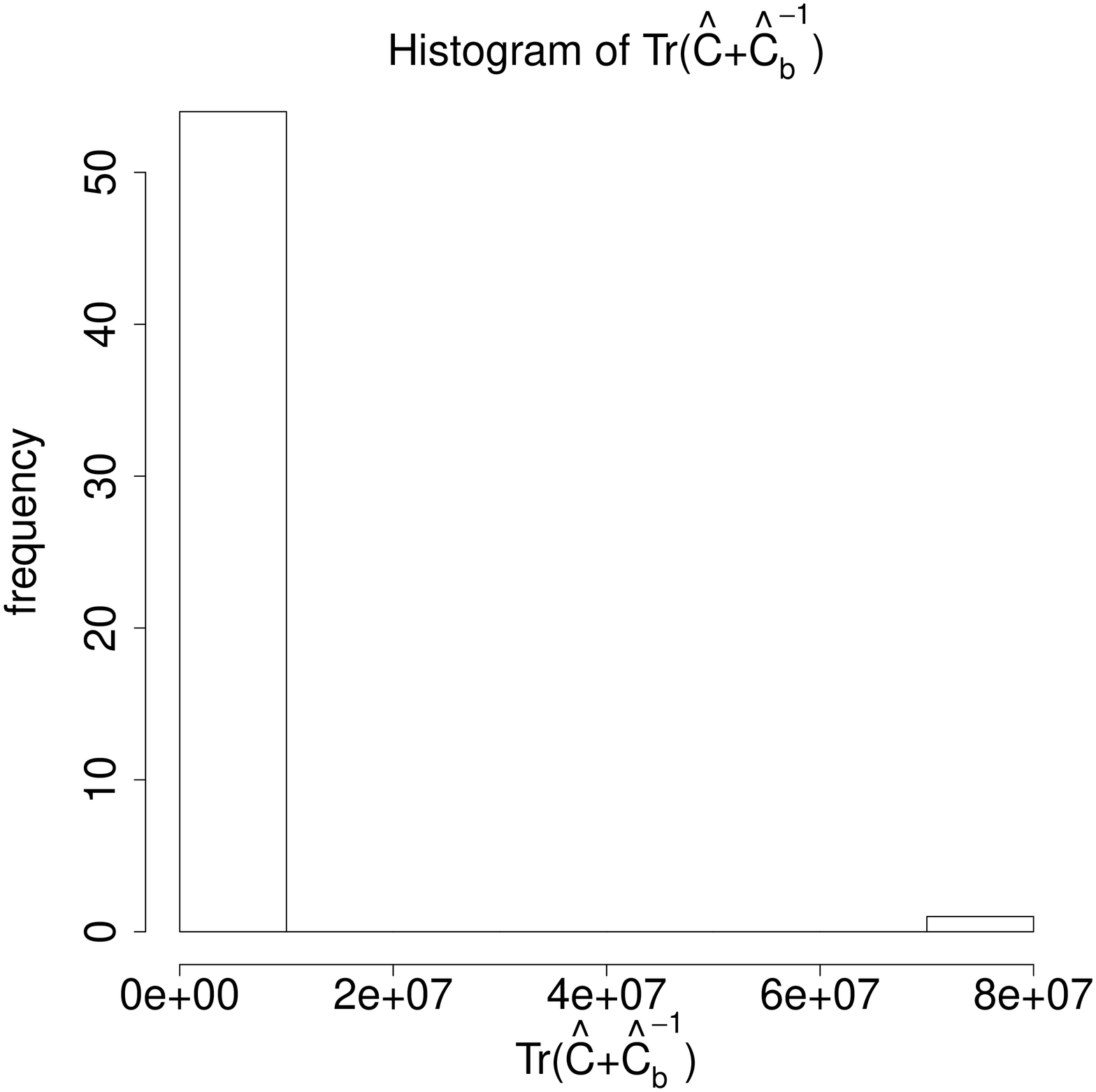}
 \end{minipage}
\caption{Histogram of the quantity
 $\trace\left(\widehat{C}_+\widehat{C}_b^{-1}\right)$ of mDEE3 when $n=10$.}
\label{eeheatts}
\end{figure}

The results are shown in Fig. \ref{fig13}-\ref{fig14}. First, we should
mention that mDEE seemed to work poorly for the data sets ``eeheat'' and
``eecool.'' These data sets include discrete covariates taking only a
few values. Thus, there are some sub data sets $D_b$ in which the
above covariates take the exactly same value. In such cases,
$\widehat{C}_b^{-1}$ based on such $D_b$ diverges to infinity. 
To see this, we show
the histogram of $\trace(\widehat{C}_+\widehat{C}_b^{-1})$ of mDEE3 for
``eeheat'' data with $n=10$ in Fig. \ref{eeheatts}. From Fig. \ref{eeheatts}, we can
see that some values of them take extremely large values.
There are some ways to avoid this difficulty. The simplest way is to
replace the empirical mean
$\frac{1}{B+1}\sum_{b=0}^B\trace(\widehat{C}_+\widehat{C}_b^{-1})$ in
\eqnum{modified} with the median of
$\left\{\trace(\widehat{C}_+\widehat{C}_0^{-1}),\trace(\widehat{C}_+\widehat{C}_1^{-1}),\cdots,
\trace(\widehat{C}_+\widehat{C}_B^{-1})\right\}$. 
Applying this idea to mDEE3, we obtain a new criterion referred to as
rmDEE (robust mDEE). 
Each panel of ``eeheat'' and ``eecool'' in Fig. \ref{fig14} contains the 
result of rmDEE instead mDEE2.
From Fig. \ref{fig14}, we can see that rmDEE worked significantly better
than mDEE1 or mDEE3.
On real-world data, mDEE1 slightly performed better than mDEE2 or
mDEE3. However, their differences are little. 
In most cases, mDEE (or
rmDEE) dominated DEE or at least performed equally. Remarkably, mDEE
always dominated ADJ except `eeheat' when $n=20$.
As a whole, mDEE (or rmDEE) often performed
the best or the second best. 
\section{Conclusion}\label{conclusion}
Even though the idea of DEE seems to
be promising, it was reported that DEE performs worse than ADJ which was the
state-of-the-art. By checking the derivation of DEE, we found that the
resultant form of DEE is valid in a sense but its derivation includes
an inappropriate part. By refining the derivation in the generalized
setting, we defined a class of valid risk estimators based on the idea
of DEE and showed that more reasonable risk estimators could be found in that class.

Both DEE and mDEE assume that a large set of unlabeled data is
available. Even though these unlabeled data can also be used to estimate
the parameter (\thatis, semi-supervised learning), 
DEE and mDEE do not use them for parameter estimation. Hence, combining
the idea of DEE with semi-supervised estimator is an interesting future
work. However, it seems not to be an easy task because the derivation of
DEE strongly depends on the explicit form of LSE. 

\section*{Acknowledgements}
This work was partially supported by JSPS KAKENHI Grant Numbers
 (19300051), (21700308), (25870503), (24500018). We thank the anonymous
 reviewers for useful comments. Some theoretical results were motivated by their comments.


\begin{thebibliography}{11}
\expandafter\ifx\csname natexlab\endcsname\relax\def\natexlab#1{#1}\fi
\expandafter\ifx\csname url\endcsname\relax
  \def\url#1{\texttt{#1}}\fi
\expandafter\ifx\csname urlprefix\endcsname\relax\def\urlprefix{URL }\fi

\bibitem[{Akaike(1970)}]{akaike70}
Akaike, H., 1970. Statistical predictor identification. {\it Annals of
  Institute Statistical Mathematics} 22, 202--217.

\bibitem[{Akaike(1972)}]{akaike72}
Akaike, H., 1972. Information theory and an extension of the maximum likelihood
  principle. {\it Second international Symposium on Information Theory},
  267--281.

\bibitem[{Bache and Lichman(2013)}]{bachelichman13}
Bache, K., Lichman, M., 2013. {UCI} machine learning repository.
\newline\urlprefix\url{http://archive.ics.uci.edu/ml}

\bibitem[{Chapelle et~al.(2002)Chapelle, Vapnik, and Bengio}]{chapelleetal02}
Chapelle, O., Vapnik, V., Bengio, Y., 2002. Model selection for small sample
  regression. {\it Machine Learning} 48, 9--23.

\bibitem[{Hastie et~al.(2001)Hastie, Tibshirani, and Friedman}]{hastieetal01}
Hastie, T., Tibshirani, R., Friedman, J., 2001. The Elements of Statistical
  Learning: Data Mining, Inference and Prediction. Springer-Verlag.

\bibitem[{Horn and Johnson(1985)}]{hornjohnson85}
Horn, R., Johnson, C., 1985. Matrix analysis. Cambridge: Cambridge University
  Press.

\bibitem[{Kawakita et~al.(2010)Kawakita, Oie, and Takeuchi}]{kawakitaetal10}
Kawakita, M., Oie, Y., Takeuchi, J., 2010. A note on small sample regression.
  {\it Proceedings of 2010 International Symposiumu on Information Theory and
  Its Applications}, 112--117.

\bibitem[{Konishi and Kitagawa(1996)}]{konishikitagawa96}
Konishi, S., Kitagawa, G., 1996. Generalized information criteria in model
  selection. {\it Biometrika} 83~(4), 875--890.

\bibitem[{Schuurmans(1997)}]{schuurmans97}
Schuurmans, D., 1997. A new metric-based approach to model selection. {\it
  Proceedings of the Fourteenth National Conference on Artificial
  Intelligence}, 552--558.

\bibitem[{Schwartz(1979)}]{schwartz79}
Schwartz, G., 1979. Estimating the dimension of a model. {\it Annals of
  Statistics} 6, 461--464.

\bibitem[{Sugiura(1978)}]{sugiura78}
Sugiura, N., 1978. Further analysts of the data by akaike' s information
  criterion and the finite corrections. {\it Communications in Statistics -
  Theory and Methods} 7~(1), 13--26.

\end{thebibliography}

\end{document}